\documentclass[epj]{svjour} 
\usepackage{amsmath}
\usepackage{graphicx}
\usepackage{epsfig}
\usepackage{feynmf}
\usepackage[dvips]{color}

\begin{document}
\abstract{Experimental prospects for studying high-energy photon-photon and photon-proton interactions at the CERN Large Hadron Collider (\textsc{lhc}) are discussed. Cross sections are calculated for many electroweak and beyond the Standard Model processes. Selection strategies based on photon interaction tagging techniques are studied. Assuming a typical \textsc{lhc} multipurpose detector, various signals and their irreducible backgrounds are presented after applying acceptance cuts. Prospects are discussed for the Higgs boson search, detection of supersymmetric particles and of anomalous quartic gauge couplings, as well as for the top quark physics.}

\title{High energy photon interactions at the LHC}
\author{J.~de~Favereau~de~Jeneret \and V.~Lema\^itre  \and Y.~Liu\thanks{\emph{Present address: }University of Science and Technology of China, Hefei, China} \and S.~Ovyn \and T.~Pierzcha\l a \and K.~Piotrzkowski \and X.~Rouby\thanks{\emph{Present address: }Physikalisches Institut, Albert-Ludwigs-Universit\"at, Freiburg, Germany} \and N.~Schul \and M.~Vander~Donckt\thanks{\emph{Present address: }IPNL, Universit\'e de Lyon, Universit\'e Lyon 1, CNRS/IN2P3, 4 rue E. Fermi 69622 Villeurbanne cedex, France} }
\institute{Universit\'e catholique de Louvain, Center for Particle Physics and Phenomenolgy (CP3), 1348 Louvain-la-Neuve, Belgium}
\date{\today}
\mail{krzysztof.piotrzkowski@uclouvain.be}
\maketitle
\section{Introduction}
\label{intro}
Photon interactions have been extensively studied at \sloppy \textsc{hera} and \textsc{lep} colliders, in processes involving exchange of quasi-real photons collinear to the incoming electron. In most of these studies the hadronic sector of photon physics has been investigated at rather limited energies. In a similar manner, a significant fraction of proton-proton collisions at the \textsc{lhc} will involve (quasi-real) photon interactions, however this time occurring at energies beyond the electroweak energy scale. Hence, the \textsc{lhc} can to some extend be considered as a high-energy photon-photon or photon-proton collider.\\

This offers a unique possibility for a novel and complementary research where the much smaller available effective luminosity is compensated by better known initial conditions and usually simpler final states. It should be noted, that this can be done parasitically to the nominal $pp$ studies. This is also, in a way, a method for approaching some of the issues to be addressed by the future lepton collider.\\

Studies of photon interactions are possible at the \textsc{lhc}, thanks to striking experimental signatures of events involving photon exchanges such as the presence of very forward scattered protons and of large rapidity gaps (\textsc{lrg}s) in forward directions.  However, to $tag$ efficiently photon induced processes and to keep backgrounds under control, some of the studied reactions require dedicated very forward proton detectors, capable in addition to measure the exchanged photon momenta~\cite{piotr_1}. Very recently and for the first time, photon induced processes have been measured in $p\bar{p}$ collisions at Tevatron using the \textsc{lrg} signature. Both in the case of the exclusive two-photon production of lepton pairs~\cite{tev1}, and of the diffractive photoproduction of J/$\psi$ mesons~\cite{tev2}, clear signals were obtained with low backgrounds.\\


The equivalent photon approximation (\textsc{epa}) can be successfully used to describe the majority of processes involving photon exchange, provided that the amplitude of a given process can be factorized into the `universal' photon exchange part and the process dependent, photon interaction part~\cite{EPA}. The photon-photon and photon-proton cross sections, $\sigma_{\gamma\gamma}$ and $\sigma_{\gamma p}$, can then be convoluted with the photon spectra \sloppy $dN(x,Q^2)$ to obtain the $pp$ cross sections, where $x=E_\gamma/E$, $E_\gamma$ is the photon energy, $Q^2$ its virtuality, and $E$ is the beam energy: 
\[
d\sigma_{pp}=\sigma_{\gamma\gamma}~dN_1~dN_2,
\]
\[
d\sigma_{pp}=\sigma_{\gamma p}~dN.
\]
When photon cross sections are not very sensitive to the photon virtuality, one can introduce the $Q^2$-integrated relative luminosity spectrum $f_\gamma$ in term of which, photoproduction cross sections read\footnote{One should note a sign error in the relevant formula D7 in Ref.~\cite{EPA}.}: 
$$
\sigma_{pp}=\int_{x_{min}}^{1} f_{\gamma}(x)~\sigma_{\gamma p}~dx.
$$
In a similar manner one can introduce the photon-photon relative luminosity spectrum 
$L_{\gamma\gamma}$ as:
\begin{equation}
\frac{d L_{\gamma\gamma}}{dW_{\gamma\gamma}}=\int_{W_{\gamma\gamma}^{2}/s}^{1} 
2W_{\gamma\gamma}~f_{\gamma}(x)~f_{\gamma}\bigg(\frac{W_{\gamma\gamma}^{2}}{xs}\bigg)\frac{dx}{xs},
\label{eq.lumi}
\end{equation} 
where $s=4E^2$, and $W_{\gamma\gamma}$ is the two-photon center-of-mass system (c.m.s.) energy, or the invariant mass of the produced system in the $\gamma\gamma\rightarrow X$ process. The two-photon cross sections can then be written as:
$$
\sigma_{pp}=\int_{W_0}^{\sqrt{s}}\sigma_{\gamma\gamma}~
\frac{d L_{\gamma\gamma}}{dW_{\gamma\gamma}}~dW_{\gamma\gamma},
$$
In the following, cross sections will always be presented as $pp$ cross sections and will be categorized as parton-parton (or generic), photoproduction (or $\gamma p$), two-photon (or $\gamma \gamma$), or diffractive cross sections to illustrate the underlying process involved. Moreover, we will refer to elastic photoproduction or two-photon cross sections, or to elastic relative luminosity spectrum, when one in photoproduction, $pp\rightarrow pX$, or both incident protons in the two-photon production, $pp\rightarrow pXp$, remain intact.\\

For the case when, due to photon exchange, one or both incident protons dissociate into some low-mass system, one introduces the so-called $inelastic$ $\gamma \gamma$ relative luminosity spectrum. In general, such a luminosity is significant, even larger than the elastic one, but the average photon virtuality for the inelastic events is much higher. As the invariant mass of the low-mass system increases, the inelastic spectrum cannot be distinguished from the one arising from photons coupling to quarks in generic proton proton collisions. In the following sections, only elastic photoproduction and two-photon processes are considered.\\ 

\begin{figure}[!ht]
\begin{center}
\epsfig{file=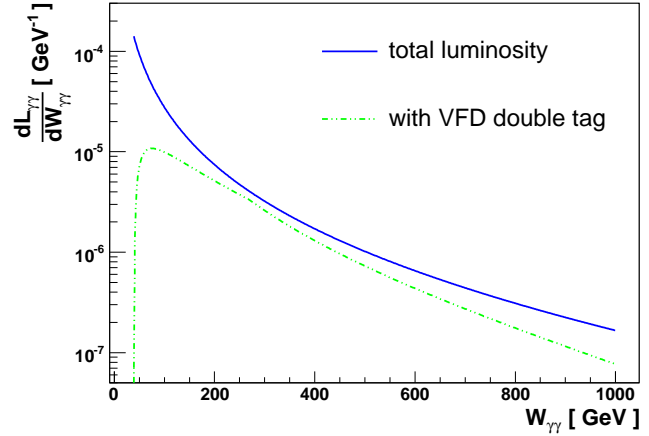 ,width=\linewidth,clip}
\caption{Elastic relative luminosity spectrum of photon-photon collisions at the \textsc{lhc}
for the integration range $Q_{min}^2<Q^2<2~\textrm{GeV}^2$ (plain blue line).
 The dashed (green) line 
shows the corresponding luminosity if energy of each photon is restricted to the tagging range 
$20~\textrm{GeV}~<~E_{\gamma}~<~900~\textrm{GeV}$.}
\end{center}
\label{lumi}
\end{figure}

As an example, the elastic relative luminosity spectrum for two-photon production can be introduced as the convolution ($d L_{\gamma\gamma} / dW_{\gamma\gamma}$ in Eq.~\ref{eq.lumi}) of the photon spectra $dN_1$ and $dN_2$ over the photon energies and virtualities for a fixed $W_{\gamma \gamma}$. It is shown in Fig.~\ref{lumi} assuming the $Q_{min}^2<Q^2< 2~\textrm{GeV}^2$ range, where $Q_{min}^2$ for low $x$ is approximately equal to 0.88$\times x^2~[\textrm{GeV}^2]$, and the upper limit is motivated by strong suppression of photon exchanges for $Q^2>1~\textrm{GeV}^2$ due to the proton electromagnetic form-factor~\cite{EPA}. Since the average photon virtualities are very low, one can safely treat all such processes as quasi-real photon interactions. The luminosity spectrum peaks at low $W_{\gamma \gamma}$, but extends far, even beyond 1 TeV (see Fig.~\ref{lumi}).\\

One can also integrate the photon relative luminosity spectrum above some initial energy $W_0$ to obtain just a single dimensionless number for easy conversions between $\sigma_{\gamma\gamma}$ and $\sigma_{pp}$. This is particularly interesting when the photon-photon cross section is not strongly energy dependent and the physics potential of high-energy photon interactions can immediately be seen. For example, the elastic photon-photon relative luminosity, obtained by integrating the luminosity spectrum in Fig.~\ref{lumi}, reaches $1\%$ for $W_{\gamma \gamma}>23$~GeV (or for $W_0=23$~GeV) and is $0.1\%$ for $W_{\gamma \gamma}>225$~GeV. Given the very large \textsc{lhc} luminosity, this leads to significant event rates of high-energy processes with relatively small photon-photon cross sections. This is even more true for the photon-proton interactions, where both energy reach and effective luminosity are much higher than for the photon-photon case. \\

Finally, photon physics can also be studied in ion collisions at the \textsc{lhc}, where low luminosity is largely compensated (for low photon energies) by high photon fluxes due to $Z^2$ enhancement, where $Z$ is the ion charge. In this case, in general lower c.m.s. energies are available for photon collisions, but the diffractive backgrounds are usually significantly lower than in the $pp$ case~\cite{ions}. Two-photon production of electron-positron pairs was recently observed in ion collisions at \textsc{rhic} \cite{rhic}.
\subsection{Tagging and forward proton detectors}
\label{sec.rp}
Tagging techniques have to be developed to allow studies of high energy photon induced processes at the \textsc{lhc}. In this paper, two well-known tagging strategies are considered. The first requires measuring the very forward scattered protons in dedicated proton detectors (Sec.~\ref{GG}) and the second, the presence of large rapidity gaps in the central detector (Sec.~\ref{GP}).\\

During the phase of low luminosity at the \textsc{lhc} (i.e. when the proton luminosity is significantly lower than $10^{33}$ cm$^{-2}$s$^{-1}$) the probability of multiple proton collisions in a single beam  crossing (or, of the so-called \textit{event pile-up}) is low. In this case, one can suppress generic $pp$ interactions by requiring \textsc{lrg}s defined as large regions in pseudorapidity, next to the centrally produced system $X$, devoid of outgoing particles. The advantage of this tagging technique is that it can be done using the central detectors only. In this case, both the elastic and inelastic photon interactions are selected, leading to higher photon luminosities but also to larger uncertainties due to less theoretically controlled inelastic photon spectra. Moreover, momenta of the exchanged photons are not measured leading to a less constrained event kinematics. The total integrated \textsc{lhc} luminosity for such no-pileup conditions cannot be precisely estimated but 1~fb$^{-1}$ seems to be a realistic guess. \\

At high luminosity, the \textsc{lrg} technique (making use of forward calorimeters) cannot be used because of large event pile-up. Therefore the use of dedicated \textit{very forward detectors} (\textsc{vfd}s) becomes mandatory. In elastic processes the small virtuality of the exchanged photons leads to very small proton scattering angles, so that the outgoing proton trajectories are determined mostly by the proton energy loss~\cite{piotr_1,hector}. Indeed, the \textsc{lhc} beamlines near the interaction points (\textsc{ip}s) include very strong dipole magnets used to separate the two proton beams before and after their crossing. These magnets will therefore deflect stronger the protons which lost significant energy. One can therefore build forward proton spectrometers by installing position sensitive detectors at large distance from the \textsc{ip} and using the \textsc{lhc} beamlines as analyzing magnets. \\ 

Two sets of such \textsc{vfd}s are assumed. The first set contains two \textsc{vfd}s located at about $220$ m (or 240~m) from the \textsc{ip} in each direction, with the inner edge of their sensors only $2$ mm from the nominal beam position in the horizontal plane. A high acceptance window is then expected for proton energy loss between $120$ to $900$~GeV. The second set of detectors will be installed at about $420$ m from the \textsc{ip}, with the detector edge at about 4 mm from the beam (also in the horizontal plane), leading to a high acceptance for \textit{tagged} photon energies between $20$ and $120$~GeV. One expects to obtain very good photon energy resolution of about $2$~GeV for low energies, and of $5$--$7$~GeV for the highest photon energies (or equivalently, of a 1\% relative energy resolution)~\cite{hector}. As demonstrated in Fig.~\ref{lumi}, such \textsc{vfd}s will allow for efficient tagging the two-photon production in the full energy range of interest. One should note that the technical proposal for such very forward detectors capable to run at high luminosity have been published~\cite{fp420} and an integrated luminosity of 100~fb$^{-1}$ can be usually assumed in this case.
\subsection{Survival probabilities}
\label{sec.Survival}
The hadronic nature of incident protons has been neglected in the above discussion of the photon induced processes at the \textsc{lhc}. However, strong interactions between the colliding protons have to be taken into account. This is usually done by introducing a correction factor to the cross section, called the \textit{survival probability}~\cite{khoze}. In the elastic case, it basically corresponds to the probability of the scattered protons not to dissociate due to the secondary soft interactions (\textit{rescattering}). Calculations of the survival factors are usually done in the impact parameter space, assuming the factorization as in the \textsc{epa}. The proton impact parameter is approximately inversely proportional to $\sqrt{Q^2}$ and is usually much bigger than the range of strong interactions. The proton survival probability is therefore generally large, close to $100\%$. Since the average $Q^2$ increases with photon energy, in general, one expects lower survival probability for higher c.m.s. energy of photon interactions. Finally, the survival probability for the two-photon interactions (with the presence of two large impact parameters associated to both protons) is usually higher than for the photon-proton processes.\\

In the inelastic case, the typical impact parameters are much smaller due to the much larger average virtuality of the exchanged photons. This usually leads to significantly smaller survival probabilities, which additionally have a different meaning. In this case, the proton has already dissociated due to the photon exchange and the main result of secondary soft interactions is an additional production of soft particles, which can populate the \textsc{lrg}s and spoil the signature of photon induced events. The calculation here is in general more complex, since it requires applying the same conditions as those used to define the \textsc{lrg}s, which in principle can be also process dependent. In summary, the \textsc{lrg} survival probability for inelastic case is not well known, leading to increased uncertainty of the evaluated photon cross sections.\\

Several studies of the survival probabilities have been done, including also the diffractive processes, where much lower values were obtained, due to small impact parameters involved. For instance, the survival factor expected at the \textsc{lhc} for the central exclusive Higgs boson production is only $\simeq 2\%$~\cite{SurDif}, whereas the one corresponding to $pp\rightarrow (\gamma\gamma \rightarrow H) \rightarrow pHp$ is $86\%$~\cite{SurGG}, and is almost $100\%$ for the elastic two-photon production of lepton pairs with low invariant mass. In photoproduction, the proton survival probability should be still high. For example, it is expected to be of about 70\% for the single $W$ boson photoproduction~\cite{SurG2}. It was also proposed that the survival probability calculations could be tested at the \textsc{lhc} by studying the exclusive production the lepton pairs~\cite{Bjorken}, and the single $W$ boson photoproduction~\cite{SurG2}.\\

The cross sections given in this paper do not include the survival probability.\\
\subsection{Background}
\label{sec.bkg.gg}

In addition to the irreducible photon-induced backgrounds, relevant for the studied process and discussed in the following sections, the diffractive processes and generic $pp$ interactions should be also considered.\\

The topology of photon interactions -- some centrally produced particles and one or two very forward scattered protons -- can also occur in diffractive processes. This diffractive background will therefore be selected along with the photon induced events. However, the transverse momentum of the diffractively scattered protons is on average significantly larger than for the photon induced events. This feature could be used to separate the two contributions on a statistical basis~\cite{piotr_1}. In addition, diffractive interactions usually result in higher multiplicities of final state particle (due to the presence of the so-called Pomeron remnants). This is also the case for the so-called resolved photon contribution (shortly discussed in see Sec.~\ref{GP}). Therefore, some dedicated exclusivity conditions can be in principle applied to control the diffractive background and to enhance the photon signals.\\

In case of photoproduction, single diffraction will result in similar topologies and to estimate the order of magnitude of its contribution one can combine the results obtained at Tevatron with the calculations of the corresponding survival probabilities at the \textsc{lhc}~\cite{SurG}. For example, \textsc{lrg} was observed in about 1\% of events with the singly produced $W$ bosons at Tevatron~\cite{W-tev}. Due to the survival probability decreasing with energy, this fraction is expected to be about 0.5\% at the \textsc{lhc}~\cite{fp420}. Similarly, the diffractive top quark pair production at the \textsc{lhc} is expected to be below 0.5\% of the total inclusive production. In this case, the diffractive production will have similar rate as for photoproduction. In contrast, for processes where the photoproduction is relatively enhanced with respect to the inclusive case (as for the associated $WH$ production, for example), the diffractive background is expected to be small. Finally, one should note that the photoproduction drops slower with energy, which favors studies of final states with large invariant mass. However, the selection of photoproduction will, in general, make use of \textsc{lrg} for which, sometimes, non-diffractive processes can also contribute. This is therefore discussed in more detail for the most affected processes in Sec.~\ref{sec.GPBack}.\\

All two-photon processes considered in this paper involve final states containing only two non-strongly interacting particles so that the diffractive backgrounds can be safely neglected. For example, the gluon 
mediated exclusive production of $W$ boson pairs ($pp\rightarrow pWWp$), is expected to be about 100 times smaller than the exclusive two-photon production at the \textsc{lhc}~\cite{khoze2}. In these studies, the double tagging using \textsc{vfd}s is assumed, so that the non-diffractive background can be also omitted.\\

At high luminosity, the event pile-up will give rise to reducible backgrounds due to accidental coincidences of detecting forward protons from diffractive processes and a generic $pp$ interaction in the central detector. Indeed, according to~\cite{cms-totem}, roughly 2 and 6\% of \textit{minimum bias} events result in a proton (mostly from single diffraction) detection by the \textsc{vfd}, at 420 and 220 m from the  \textsc{ip}, respectively. These backgrounds can be suppressed by using exclusivity conditions, kinematics and timing constraints~\cite{fp420}. Its full analysis will be the subject of future publications. Since discussions in this paper usually assume low luminosity conditions at the \textsc{lhc}, these backgrounds are neglected in the following analyses.

\section{Two-photon production}
\label{GG}
\subsection{Introduction}
\label{sec.GGIntro}
\label{sec.GG}
Elastic two-photon processes yield very clean event topologies at the \textsc{lhc}: two very forward protons measured far away from the \textsc{ip} and a few centrally produced particles. In addition, the photon momenta can be precisely measured using the \textsc{vfd}s, allowing to reconstruct the total center of mass energy of the photon collisions irrespective of any missing longitudinal degrees of freedom in the final state. Besides, if only non-strongly interacting particles are produced the large backgrounds from parton-parton interactions at the \textsc{lhc} can be sufficiently suppressed.\\

\begin{figure}[!ht]  
\begin{center}
\epsfig{file=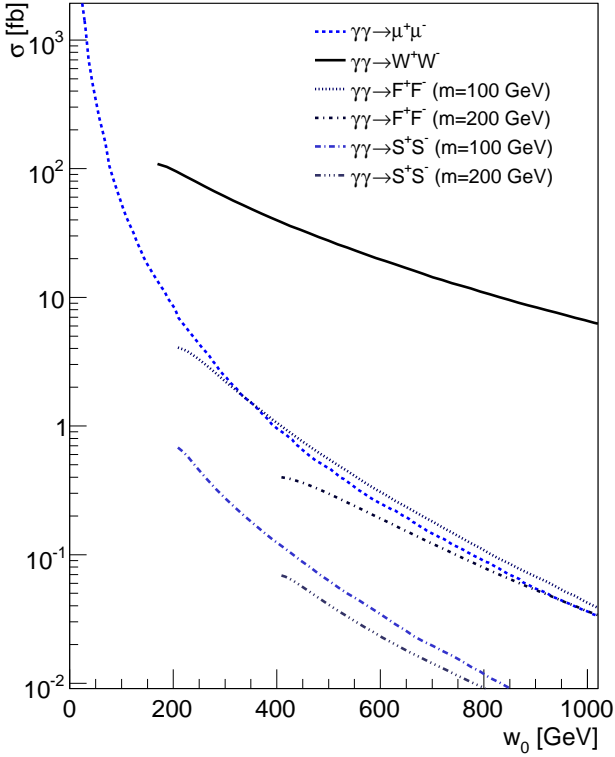,width=\linewidth} 
\caption{\small{Cross sections for various two-photon processes, $pp\rightarrow pXp$, as a function
 of the minimal photon-photon c.m.s. energy $W_0$. The $pp \rightarrow p\mu^+\mu^-p$ cross section has been computed for $p_T^{\mu}> 2$~GeV and $|\eta^\mu|<$3.1. No cut is applied for the other processes.}}
\label{IntGG} 
\end{center}
\end{figure}

To illustrate the physics potential of two-photon interactions, some cross sections have been computed using the modified\footnote{First public \textsc{mg/me} release containing the photon interactions based on the \textsc{epa} are available starting from version V4.0.5.} MadGraph/MadEvent (\textsc{mg/me})~\cite{mad} and are shown in Fig.~\ref{IntGG} for several $pp\rightarrow(\gamma \gamma \rightarrow X)\rightarrow pXp$ processes.\\

Clearly, several two-photon processes have sufficiently large cross sections at the \textsc{lhc} and should therefore allow for interesting studies. The existence of new massive charged particles could be verified via the two-photon pair production -- in particular, pairs of supersymmetric particles have been considered~\cite{zerwas}. Similarly, the two-photon exclusive production of the Standard Model (\textsc{sm}) Higgs boson has been studied~\cite{higgs,piotr_1}. This process has a too small cross section of $0.154$ fb to become a discovery channel, but could become interesting in case of an enhanced $H\gamma\gamma$ coupling. Moreover, the two-photon exclusive production of lepton pairs can be used for an absolute luminosity normalization at the \textsc{lhc}~\cite{lumi}.\\

Cross sections for several photon induced processes are summarized in Tab.~\ref{backGG}. Since the cross sections for pair production depend only on charge, spin and mass of the produced particles, the results are shown for singly charged and colorless fermions and scalars of two different masses.\\
\begin{table}[!ht]
\begin{center}
\begin{tabular}{ll||c |c}
\hline
\multicolumn{2}{l||}{Processes}  & $\sigma$ [fb] & Generator  \\\hline
$\gamma \gamma \rightarrow$ & $\mu^+ \mu^-$ &  72 500 & \textsc{lpair}~\cite{lpair}\\
& $W^+ W^-$ & 108.5 & \textsc{mg/me} \\ 
& $F^+ F^-$ (m = 100~GeV) & 4.064 & $\scriptscriptstyle{//}$ \\
& $F^+ F^-$ (m = 200~GeV) & 0.399 & $\scriptscriptstyle{//}$ \\
& $S^+ S^-$ (m = 100~GeV) & 0.680 & $\scriptscriptstyle{//}$ \\
& $S^+ S^-$ (m = 200~GeV) & 0.069 & $\scriptscriptstyle{//}$ \\
& $H \rightarrow b\overline{b}$ (m = 120~GeV) & 0.154 & $\scriptscriptstyle{//}$ \\
\hline
\end{tabular}
\end{center}
\caption{\small{Cross sections for several two-photon processes $pp \rightarrow pXp$  ($F$ stands for fermion, $S$ for scalar), obtained using two generators. The $pp \rightarrow p\mu^+\mu^-p$ cross section has been computed for $p_T^{\mu}>$ 2~GeV and $|\eta^\mu|<$3.1.}}
\label{backGG}
\end{table}

Realistic studies, including acceptance cuts, of the interesting $\gamma \gamma$  processes at the \textsc{lhc} are presented in the following sub-sections. All generated processes were hadronised using the modified \textsc{pythia} generator\footnote{First public \textsc{pythia} release including the photon interactions based on the \textsc{epa} is available starting from version $6.412$.}~\cite{pythia}. 

\subsection{Muon pairs}
\label{sec.muonpairs}
Two-photon exclusive production of muon pairs at the \textsc{lhc} has a well known cross section, and requires very small hadronic corrections~\cite{durham}. Its large cross section, reaching $\sigma = 72.5$~pb for muons at significant transverse momenta (see Tab.~\ref{backGG}), and the small theoretical uncertainty makes it a perfect candidate for the \textsc{lhc} absolute luminosity measurement. Thanks to its striking signature the selection procedure is very simple. It just requires two opposite charge muons within the central detector acceptance ($|\eta|<2.5$), at transverse momenta above some not too high threshold ($p_T^{\mu}>3$ or $10$~GeV). The cross sections after acceptance cuts ($\sigma_\textrm{acc}$), with or without the requirement of at least one \textsc{vfd} tag, are presented in Tab.~\ref{tab.lpaircuts}.\\ 

\renewcommand{\arraystretch}{1.3}
\begin{table}[!ht]
\begin{center}
\begin{tabular}{c||c|c}
      \hline  Cross section [fb]  & $p_T^{\mu} > 3$~GeV & $p_T^{\mu} > 10$~GeV \\ \hline
        $\sigma_{acc}$           & $21~600$  & $1~340$  \\
        $\sigma_{acc}$ (with {\sc vfd} tag) & $7~260$  & $1~270$  \\
        \hline
\end{tabular}
\end{center}
\caption{Cross section for $pp \rightarrow p\mu^+\mu^- p$ after applying acceptance cuts defined by $|\eta^\mu|<2.5$ and two lower values for $p_T^{\mu}$, with and without \textsc{vfd} tagging.}
\label{tab.lpaircuts}
\end{table}
\renewcommand{\arraystretch}{1.}
As a result, about $400$ muon pairs should be detected in a $12$~h long run at the average luminosity of $5\times10^{32}~\textrm{cm}^{-2} \textrm{s}^{-1}$. Another interesting application of these exclusive dimuon events is the possibility to perform the absolute calibration of the very forward detectors. Indeed, since the momentum of the produced muons is expected to be very well measured by the central tracking detector, the forward proton energy can be precisely predicted using the kinematics constrains. This would, in turn, provide a very good control of the photon energy reconstruction~\cite{hector}. A precise re-calibrations of the \textsc{vfd}s in case of misalignment of the \textsc{lhc} beamline elements would also be possible. Furthermore, the large cross section could even allow to perform run-by-run calibrations, as the requirement of at least one \textsc{vfd} tag results in more than $150$ events per run~\cite{rouby}.
\subsection{$W$ and $Z$ boson pairs}
\label{sec.WWZZ}
%
%
\renewcommand{\arraystretch}{1.3}

The two-photon exclusive production of $WW$ and $ZZ$ pairs is particularly well suited for studies of anomalous quartic gauge couplings (\textsc{aqgc}s) $\gamma\gamma WW$ and $\gamma\gamma ZZ$. Such anomalous couplings appear in many extensions of the \textsc{sm}, and in particular could directly reveal the exchange of new, heavy bosons \cite{Belanger:1992qh,wudka}.\\

In the \textsc{sm}, the two-photon cross section of exclusive $W$ pair production, $pp \rightarrow pW^+W^-p$, is large, about $108$~fb$^{-1}$. The topology resulting from fully muonic decay of both $W$ bosons still has a sizable cross section after applying basic acceptance cuts for both muons, as shown in Tab.~\ref{tab.wwcut}. Interestingly, cross sections after acceptance cuts and requiring at least one proton to be detected by the \textsc{vfd}s reduce it only slightly, and render this process basically background free. Therefore, this reaction at the \textsc{lhc} could provide a sensitive test-ground for the \textsc{sm} quartic gauge boson couplings. In contrast, the two-photon cross section of exclusive $Z$ pair production, $pp \rightarrow pZZp$,
is very small in the \textsc{sm} since this process is not allowed at the tree level. The observation of a few events of this type could therefore reveal the existence of the \textsc{aqgc}s, hence of physics beyond the \textsc{sm}. 

\begin{table}[!ht]
\begin{center}
\begin{tabular}{c||c|c}
        \hline Cross section [fb]  & $p_T^{\mu} > 3$~GeV & $p_T^{\mu} > 10$~GeV \\ \hline
        $\sigma_{acc}$           & $0.80$  & $0.76$  \\
        $\sigma_{acc}$ (with \textsc{vfd} tag) & $0.70$  & $0.66$  \\
        \hline
\end{tabular}
\end{center}
\caption{Cross section for $pp\rightarrow (\gamma \gamma \rightarrow W^+ W^-) \rightarrow p\mu^+ \mu^- \bar{\nu_{\mu}}\nu_{\mu}p$ after applying acceptance cuts defined by $|\eta^\mu|<2.5$ and two lower values for $p_T^{\mu}$, with and without \textsc{vfd} tagging.}
\label{tab.wwcut}
\end{table}

Anomalous couplings can be introduced by building an effective lagrangian which models low energy behavior of a wide class of possible extensions of the \textsc{sm}. In the present analysis, we introduce the so-called genuine \textsc{aqgc}s which do not require associated trilinear gauge boson couplings. This is realized by adding new interaction terms of dimension six. In order to be consistent with precise electroweak data, the simplest effective lagrangian has to conserve local $\mathrm{U}(1)_{em}$ and custodial SU(2)$_c$ symmetries. Then, by imposing conservation of discrete C and P symmetries one finally obtains four new terms \cite{Belanger:1992qh}:
\begin{eqnarray}
\mathcal{L}_6^0 &~~=~~&  - \frac{e^2}{8} {\bf\frac{a_0^{\mathrm{W}}}{\Lambda^2}} F_{\mu\nu}F^{\mu\nu} {W}^{+\alpha}{W}^-_{\alpha}\nonumber\\
& & - \frac{e^2}{16\cos^2\theta_W} {\bf\frac{a_0^{\mathrm{Z}}}{\Lambda^2}} F_{\mu\nu}F^{\mu\nu} {Z}^{\alpha}{Z}_{\alpha},
 \nonumber \\
\mathcal{L}_6^{\mathrm{c}} &~~=~~&  - \frac{e^2}{16} {\bf\frac{a_{\mathrm{c}}^{\mathrm{W}}}{\Lambda^2}} F_{\mu\alpha}F^{\mu\beta} ({W}^{+\alpha}{W}^-_{\beta}+
W^{-\alpha}W^+_{\beta}) \nonumber\\
& & - \frac{e^2}{16\cos^2\theta_W} {\bf\frac{a_{\mathrm{c}}^{\mathrm{Z}}}{\Lambda^2}} F_{\mu\alpha}F^{\mu\beta} {Z}^{\alpha}{Z}_{\beta}.
\label{eq.AnomalLagrangian}
\end{eqnarray}
where $a_0^{\mathrm{W}}, a_0^{\mathrm{Z}},a_c^{\mathrm{W}}, a_c^{\mathrm{Z}}$ are four anomalous couplings, $F^{\mu\nu}$ is the electromagnetic field strength tensor, $W^+, W^-,Z$ are weak gauge boson vector fields, $e$ is the electron charge, $\theta_W$ is the Weinberg angle and $\Lambda$ is the energy scale of the new physics. Since this general lagrangian has been assumed for studies at LEP \cite{OPAL.limits}, this is also assumed in this study such that it can be directly compared with existing measurements.

Using this formalism, the $WW$ and $ZZ$ cross sections can be written as a function of the anomalous parameters:
\begin{equation}
\sigma = \sigma_{SM} + \sigma_0 a_0 + \sigma_{00} a^2_0 + \sigma_{c} a_c
+ \sigma_{cc} a^2_c + \sigma_{0c} a_0 a_c .
\label{ex.anom}
\end{equation}
For a fixed cross section $\sigma$, defined after acceptance cuts or not, this corresponds to an ellipse in the $a_0$, $a_c$ plane. All anomalous couplings equal to zero within the \textsc{sm}.\\

The sensitivity to \textsc{aqgc}s has been investigated using the di-lepton signature: two leptons ($\ell=e$ or $\mu$) within the acceptance cuts defined by $|\eta|<2.5$ and $p_T>10$~GeV. For the $WW$ production, it corresponds to the fully-leptonic subprocess $\gamma \gamma \rightarrow W^+W^- \rightarrow \ell^+ \ell^-\nu\bar{\nu}$ while for the $ZZ$ production, to the semi-leptonic subprocesses $\gamma\gamma\rightarrow ZZ\rightarrow \ell^+ \ell^- j j$, where $j$ refers to jets. It is assumed that the \textsc{sm} two-photon production of $WW$ is the only background source for this particular channel. 

Under this assumption, the expected $95\%$ confidence level (CL) upper limits of the number of events $\lambda^{\textrm{up}}$ is calculated\footnote{This is a simplified approach and for more precise analysis one should use formula from \cite{Zech:1988un}, however the difference in the obtained limits is small, less than 20\%.} assuming the number of observed events $N_\textrm{obs}$ equal to the \textsc{sm} prediction, for a given luminosity $L$.\\

\begin{equation}
\sum^{N_\mathrm{obs}}_{k=0} P(\lambda^\mathrm{up} = \sigma ^\mathrm{up}~L; k) = 1 - \mathrm{CL}.
\label{eq.simple-Poisson}
\end{equation}
Equivalently the $95\%$ CL upper limits on the observed cross section $\mathrm{\sigma^{up}}$ has been calculated for $L=1$~fb$^{-1}$ and $L=10$~fb$^{-1}$, and is shown in Tab.~\ref{tab.xs-limits}.

\begin{table}[!ht]
\begin{center}
\begin{tabular}{l||c|c}
\hline
Cross section $\sigma^\textrm{up}_\textrm{acc}$~[fb] & \multicolumn{2}{c}{Limits}  \\
                            & $L=1~\textrm{fb}^{-1}$ & $L=10~\textrm{fb}^{-1}$ \\
\hline
$\gamma\gamma\rightarrow W^+W^-$ ($\mathrm{\sigma^{\textsc{sm}}_{acc}=4.1~fb}$) & 9.2 & 5.3  \\
$\gamma\gamma\rightarrow ZZ$  ($N_\mathrm{obs} = 0$)                        & 3.0 & 0.30  \\ \hline
\end{tabular}
\end{center}
\caption{Expected 95$\%$ CL upper limits for the cross sections after acceptance cuts for subprocesses $\gamma\gamma\rightarrow W^+W^-\rightarrow l^+l^-\nu\bar{\nu}$ and $\gamma\gamma\rightarrow ZZ\rightarrow l^+l^- j j$. The assumed numbers of the observed $WW$ events correspond to the \textsc{sm} prediction calculated using \textsc{mg/me} and applying the acceptance cuts on leptons only, defined by $|\eta|<2.5$ and $p_T>10$~GeV. Here $l = e, \mu$ and $j$ refer to jets.
}
\label{tab.xs-limits}
\end{table}

\begin{figure}[!ht]  
\begin{center}
\epsfig{file=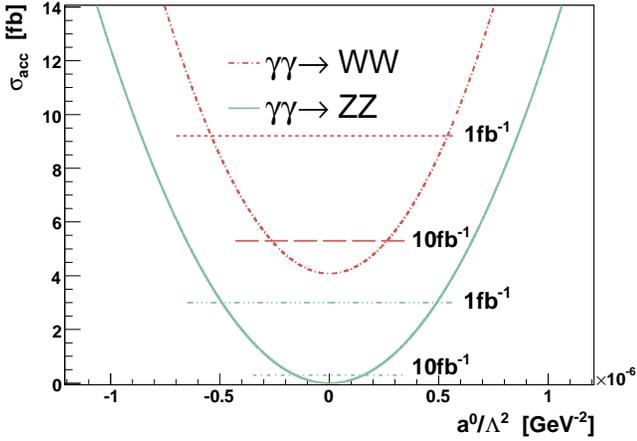,width=\linewidth,clip}
\caption{Cross section of ${pp \rightarrow (\gamma\gamma\rightarrow W^+W^-)\rightarrow p \ell^+ \ell^-\nu\bar{\nu} p}$ and ${pp \rightarrow (\gamma\gamma\rightarrow ZZ)\rightarrow p\ell^+ \ell^- j j p}$ after lepton acceptance cuts as a function of anomalous quartic vector boson coupling $\mathrm{a_0 / \Lambda^2}$ ($\mathrm{a_C / \Lambda^2=0}$) together with $95\%$ CL upper limits of the cross sections. 
}
\label{fig.sigma_aaww_aazz}
\end{center}
\end{figure}

\begin{figure}[!ht]
\begin{center}
\epsfig{file=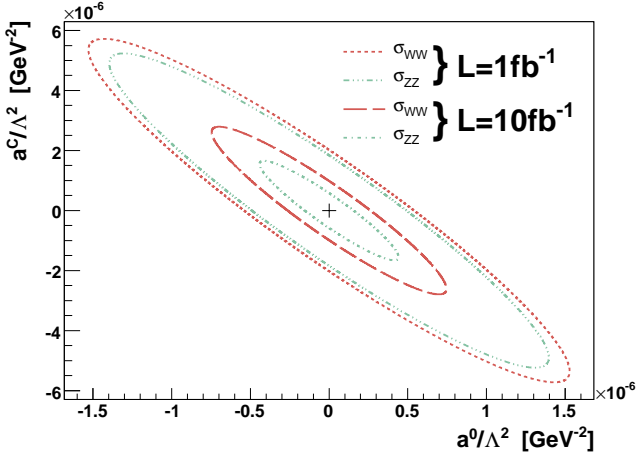,width=\linewidth,clip}
\caption{Profiles of the 95$\%$ CL upper limits of the cross section after leptons acceptance cuts for ${pp \rightarrow (\gamma\gamma\rightarrow W^+W^-) \rightarrow p\ell^+ \ell^-\nu\bar{\nu}p}$ and ${pp \rightarrow (\gamma \gamma \rightarrow ZZ) \rightarrow p\ell^+ \ell^- j j p}$ processes, as a function of relevant anomalous couplings $\mathrm{a_0 / \Lambda^2}$ and $\mathrm{a_{\mathrm{c}} / \Lambda^2}$. The contours are shown assuming two values of the integrated $pp$ luminosity $L$.
}
\label{fig.contour_aaww_aazz}
\end{center}
\end{figure}

The calculated upper limits on cross section can be easily converted to the limits on anomalous quartic couplings 
as presented in Fig.~\ref{fig.contour_aaww_aazz} where the $95\%$ CL contours are shown. In Fig.~\ref{fig.sigma_aaww_aazz} one parameter limits (the other anomalous coupling is set to zero) are shown, and the obtained limits are quoted in Tab.~\ref{tab.CL}.

\begin{table}[!ht]
\begin{center}
\begin{tabular}{c||c|c}
\hline
Coupling & \multicolumn{2}{c}{ Limits $\mathrm{[10^{-6}~\textrm{GeV}^{-2}~]}$} \\  
         &$L=1~\textrm{fb}^{-1}$ & $ L = 10~\textrm{fb}^{-1}$ \\
\hline
 $ |a^Z_0/\Lambda^2| $& $ 0.49 $ & $ 0.16 $ \\
 $ |a^Z_C/\Lambda^2| $& $ 1.84 $ & $ 0.58 $ \\
 $ |a^W_0/\Lambda^2| $& $ 0.54 $ & $ 0.27 $ \\
 $ |a^W_C/\Lambda^2| $& $ 2.02 $ & $ 0.99 $ \\
\hline
\end{tabular}
\end{center}
\caption{Expected limits for anomalous quartic vector boson couplings at $95\%$ CL for two values of integrated luminosities. Limits on a given coupling are calculated assuming a null value for other couplings.}
\label{tab.CL}
\end{table}
\renewcommand{\arraystretch}{1.}

The obtained limits are about $10,000$ times better than the best limits established at \textsc{lep2} \cite{OPAL.limits}. One should note however, that this sensitivity is driven mostly by photon-photon interactions at $W_{\gamma \gamma}>1$~TeV, giving rise to an issue of unitarity violation. It has been shown that indeed for the anomalous parameters corresponding to the above limits the unitarity violation occurs at energies above 2--3~TeV \cite{tomek}. One way of avoiding it is by introducing \textit{ad hoc} form-factors which dump the cross-sections above a chosen energy scale. This is usually done for related studies performed at hadron colliders and leads to somewhat reduced sensitivities \cite{tomek}. On the other hand, one can limit the highest accepted energies in photon-photon collisions by requiring double \textsc{vfd} tags. In this case, the sensitivity should be significantly improved by much increased kinematical constrains and possibility of using several differential distributions as a function of $W_{\gamma \gamma}$ to search for deviations from the \textsc{sm}. These preliminary results clearly show large and unique potential of such studies at the \textsc{lhc}, and their sensitivity to New Physics above 1 TeV scale.\\ 

A similar study for anomalous triple gauge couplings can be also performed. However, in this case the expected sensitivities are not as large as for the case of \textsc{aqgc}s \cite{royon}.\\

The unique signature of relatively high $p_T$ lepton with a large acoplanarity (and large missing transverse momentum) allows strong reduction of backgrounds. Thus, the two-photon production of tau-lepton pairs can be completely neglected. Moreover, the exclusive diffractive production of gauge boson pairs is also negligible and, finally, the inclusive partonic production can be also very efficiently suppressed. For example, the total parton-parton production of $W$ boson pairs assuming leptonic decays, and the lepton acceptance cuts, has cross section of about $1$~pb of at the \textsc{lhc}. Therefore, by applying either the double tag using \textsc{vfd}s, or by using the double \textsc{lrg} signature, one can reduce it very strongly.
\subsection{Supersymmetric pairs}
\label{sec.SusyBSM}
\begin{figure}[hhh]
\begin{center}
\epsfig{file=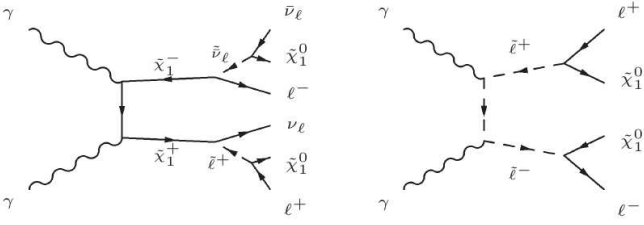,width=\linewidth}
\caption{\small{Most relevant Feynman diagrams for \textsc{susy} pair production with leptons in
 the final state. Left: chargino production and decay in a charged/neutral slepton and a
 neutral/charged fermion; Right: slepton production and decay.}}
\label{fig.diag.susy}
\end{center}
\end{figure}

The interest in the two-photon exclusive production of pairs of new charged particles is three-fold: 1. It provides a new, complementary and very simple production mechanism where the cross-sections depend strongly on the spin of the produced particles; 2. For events with double \textsc{vfd} tag very interesting constraints on the masses of new particles could be obtained; 3. Finally, in case of low mass supersymmetry (\textsc{susy}), simple final states are usually expected without complex cascade decays, resulting in very good background conditions especially for the fully leptonic final states composed of two charged leptons with large missing energy (and large lepton acoplanarity). The following discussion concentrates on such a fully leptonic \textsc{susy} case only. The corresponding Feynman diagrams are shown at Fig.~\ref{fig.diag.susy}. It should be noted that the kinematics is different for events with chargino ($\tilde{\chi}$) pairs since they decay into leptons through a short chain. Other sleptons ($\tilde{\ell}=\tilde{e},\tilde{\mu}, \tilde{\tau}$ ) usually decay directly into leptons, resulting in different transverse momentum spectra. In all the cases, two lightest supersymmetric particles (\textsc{lsp}s) escape the detection.\\

The semi-leptonic case with one lepton and one $\tau$-jet ($\tau_h$) in the final state could also be considered, but the statistical improvement only becomes relevant for the supersymmetric large $\tan\beta$ models where couplings to $\tau$ and stau ($\tilde{\tau}$) are larger than for other (s)leptons.\\ 

For this study, three low mass benchmark (namely LM1, LM2 and LM6) in parameter space of the theory with gravity mediated supersymmetry breaking, constrained by the post-\textsc{wmap} research~\cite{wmap},  have been chosen:
\begin{itemize}                                               
\item LM1, very light \textsc{lsp}, light $\tilde{\ell}$, light $\tilde{\chi}$ and $\tan\beta$=10;
\item LM2, medium \textsc{lsp}, heavy $\tilde{\ell}$, heavy $\tilde{\chi}$ and $\tan\beta$=35;    
\item LM6, heaviest \textsc{lsp}, light right $\tilde{\ell}$, heavy left $\tilde{\ell}$, heavy $\tilde{\chi}$ and $\tan\beta$=10.
\end{itemize}                                 
The masses of the supersymmetric particles were obtained by running the renormalisation group equations with SuSpect~\cite{suspect}, and are indicated in Tab.~\ref{tab.mass.susy}.\\
\begin{table}[!ht]                                                                                                   
\begin{center}
\begin{tabular}{c||c c c}                                                                                             
\hline
m [GeV] & LM1 & LM2 & LM6 \\                                                                            
\hline
 $\tilde{\chi}_1^0$ & 97 & 141 & 162 \\  
 $\tilde{\ell}_R^+$ & 118 & 229 & 175 \\
 $\tilde{\ell}_L^+$ & 184 & 301 & 283 \\                                     
 $\tilde{\tau}_1^+$ & 109 & 156 & 168 \\
 $\tilde{\tau}_2^+$ & 189 & 313 & 286 \\                                      
 $\tilde{\chi}_1^+$ & 180 & 265 & 303 \\
 $\tilde{\chi}_2^+$ & 369 & 474 & 541 \\
 $H^+$              & 386 & 431 & 589 \\
\hline
\end{tabular}
\caption{\small{Masses of \textsc{susy} particles, in GeV, for different benchmarks (here $\ell=e,\mu$.)}}
\label{tab.mass.susy}
\end{center} 
\end{table}

Event samples in Monte Carlo simulation involving the exclusive production of pair of particles predicted by \textsc{susy} and the \textsc{sm} were produced using the modified \textsc{CalcHEP} generator\footnote{First public \textsc{CalcHEP} release including the photon interactions based on the \textsc{epa} is available starting from version 2.3.6.}~\cite{Puk03}. The hadronization is performed by the modified \textsc{pythia} generator, and the following lepton acceptance cuts have been applied: $p_T >$ 3~GeV or 10~GeV, $\vert \eta \vert < 2.5$. The only irreducible background for this type of processes is the exclusive $W$ pair production. Indeed, direct lepton pairs $pp \to p\ell^+ \ell^-p$ can be suppressed using for example large acoplanarity cuts. Cross sections before and after acceptance cuts are shown in Tab.~\ref{tab.susy-sigma-a}.\\

\begin{table}[!ht]
\begin{center}
\begin{tabular}{p{0.25cm}c||c|c|c}
\hline
\multicolumn{2}{c||}{Processes}                        & $\sigma$ [fb] & \multicolumn{2}{c}{$\sigma_{acc}$ [fb]} \\
    &             &                                                & \scriptsize{$p_T > 3$~GeV} & \scriptsize{$p_T > 10$~GeV} \\
\hline
LM1 & $\gamma\gamma\to$ $\tilde{\ell}_R^+\tilde{\ell}_R^-$         & 0.798 & 0.627         & 0.475 \\
    & \ \ \ \ \ \ \ \ $\tilde{\ell}_L^+\tilde{\ell}_L^-$           & 0.183 & 0.142         & 0.133 \\
    & \ \ \ \ \ \ \ \ \ $\tilde{\tau}_i^+\tilde{\tau}_i^-$         & 0.604 & 0.023         & 0.006 \\
    & \ \ \ \ \ \ \ \ \ $\tilde{\chi}_i^+\tilde{\chi}_i^-$         & 0.642 & 0.111         & 0.035 \\
    & \ \ \ \ \ \ \ \ \ \ $H^+H^-$                                 & 0.006 & -             & - \\
\hline
LM2 & $\gamma\gamma\to$ $\tilde{\ell}_R^+\tilde{\ell}_R^-$         & 0.086 & 0.074         & 0.073 \\
    & \ \ \ \ \ \ \ \ $\tilde{\ell}_L^+\tilde{\ell}_L^-$           & 0.032 & 0.014         & 0.012 \\
    & \ \ \ \ \ \ \ \ \ $\tilde{\tau}_i^+\tilde{\tau}_i^-$         & 0.175 & 0.008         & 0.001 \\
    & \ \ \ \ \ \ \ \ \ $\tilde{\chi}_i^+\tilde{\chi}_i^-$         & 0.157 & 0.008         & 0.002 \\
    & \ \ \ \ \ \ \ \ \ \ $H^+H^-$                                 & 0.004 & -             & - \\
\hline
LM6 & $\gamma\gamma\to$ $\tilde{\ell}_R^+\tilde{\ell}_R^-$         & 0.219 & 0.176         & 0.087 \\
    & \ \ \ \ \ \ \ \ $\tilde{\ell}_L^+\tilde{\ell}_L^-$           & 0.040 & 0.036         & 0.035 \\
    & \ \ \ \ \ \ \ \ \ $\tilde{\tau}_i^+\tilde{\tau}_i^-$         & 0.146 & 0.003         & 0.001 \\
    & \ \ \ \ \ \ \ \ \ $\tilde{\chi}_i^+\tilde{\chi}_i^-$         & 0.095 & 0.035         & 0.030 \\
    & \ \ \ \ \ \ \ \ \ \ $H^+H^-$                                 & 0.001 & -             & - \\
\hline
SM \ & \ \ \ $\gamma\gamma\to$ $W^+ W^-$                            & 108.5 & 4.257         & 3.652 \\
\hline
\end{tabular}
\caption{\small{Cross sections before and after acceptance cuts, for \textsc{susy} pairs for different benchmarks, as well as for the the irreducible background. ($\tilde{\ell}_{R} = \tilde{e}_{R}$, $\tilde{\mu}_{R}$ \ \ $\tilde{\tau}_{i} = \tilde{\tau}_{1}$, $\tilde{\tau}_{2}$ \ \ $\tilde{\chi}_{i} = \tilde{\chi}_{1}$, $\tilde{\chi}_{2}$)}.}
\label{tab.susy-sigma-a}
\end{center}
\end{table}

Because $\tilde{\tau}$ pairs decay mainly into $\tau$ pairs, $\sim$ 85\% of the resulting decay products contain at least one $\tau$-jet, which due to small transverse momentum will be difficult to select. Rate of accepted di-leptonic final states from $\tilde{\tau}$ pairs is then reduce to few percents of the initial cross-section. \\

The major contribution from sleptons ($\tilde{e}, \tilde{\mu}, \tilde{\tau}$) will arise from the right- and left-handed $\tilde{e}$ and $\tilde{\mu}$ which, in general, produce central pairs of leptons with high acceptance. Selection efficiency is $\sim~80$\% for the $3$~GeV $p_{T}$ threshold, except for left $\tilde{\ell}_L$ in LM2 model where it falls down to $\sim~40$\% since $\tilde{\ell}_L^- \to \tilde{\chi}_1^- + \nu_{\ell}$ branching ratio is significant in contrast to the other cases.\\

The cross sections after acceptance cuts for pairs of charginos vary from one scenario to another mainly because of significantly different branching ratios. For instance, for LM2 benchmark, the $\tilde{\chi}_1^+ \to \tilde{\tau}_1^+ + \nu_\tau$ decay mode is dominant, producing mainly $\tau$-pairs as final states. Acceptance for di-lepton detection is then reduced to roughly $5\%$, as for the $\tilde{\tau}$ pairs in LM1 benchmark. On the contrary, decays such as $\tilde{\chi}_1^+ \to \tilde{\nu}_\ell + \ell^+$ are significant in LM1 and LM6 and increase the efficiency.\\

Exclusive production of charged Higgs boson pairs with di-leptonic final states is rather difficult to detect since the $H^+ \to \bar{b}t$ branching is dominant. The contribution to the signal is therefore lower than $10^{-3}$~fb.\\

Discovery of supersymmetry via two-photon interactions at the \textsc{lhc} is \emph{a priori} hard due to low production cross sections. Nevertheless, the very clean final states and possibilities of strong background rejection could allow for detection of pairs of light \textsc{susy} particles. For instance, the production of supersymmetric slepton pairs generally imply the appearance of two leptons with identical flavor. Selection of same flavor leptons would therefore decrease the irreducible $WW$ background by a factor of two, without affecting efficiency of the signal selection. It would, in turn, also allow a precise independent background normalization.\\

\begin{figure}[!ht]
\begin{center}
\begin{tabular}{c}
\epsfig{file=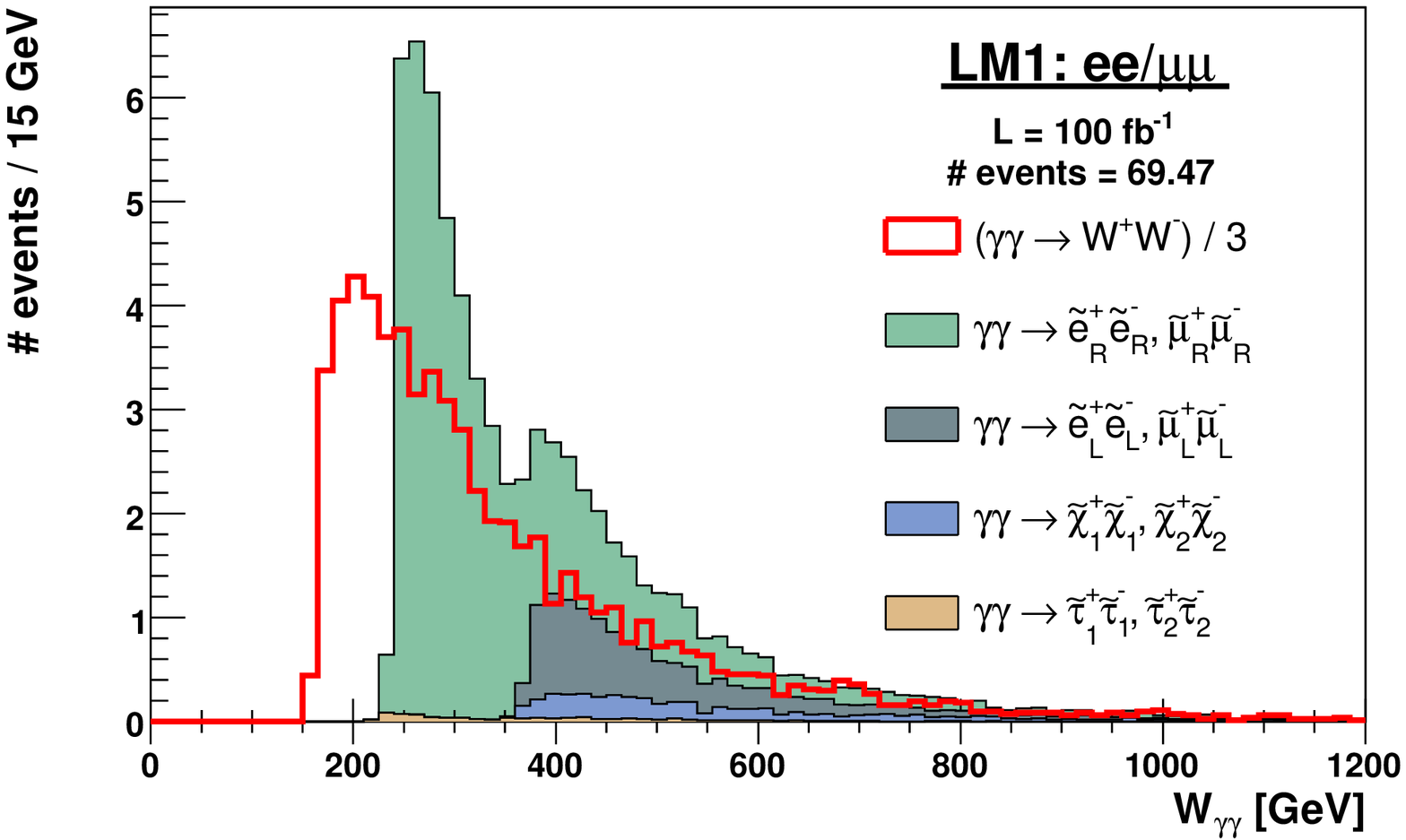, width=\linewidth,clip} \\
\epsfig{file=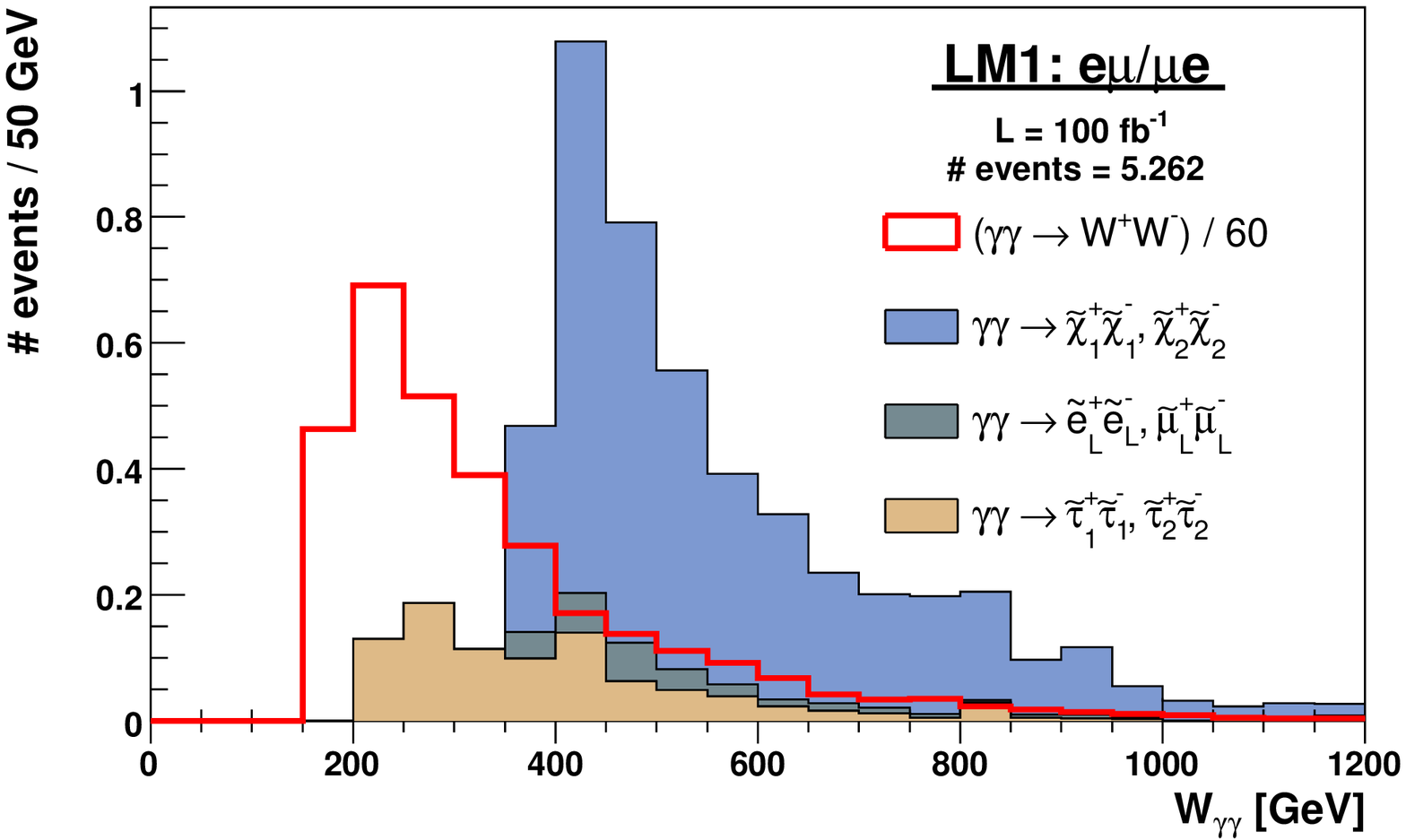, width=\linewidth,clip} 
\end{tabular}
\end{center}
\caption{\small{Expected photon-photon invariant mass ($W_{\gamma \gamma} = 2 \sqrt{E_{\gamma_1} E_{\gamma_2}} $) distributions and number of events for benchmark point LM1, after an integrated luminosity of 100~fb$^{-1}$. The event selection requires two detected protons in \textsc{vfd}s and two detected leptons with $p_T > 3$~GeV and $|\eta| <$ 2.5, with identical flavors (top histogram) or different flavor (bottom histogram). From the cumulative distribution of the signal, the various peaks could allow a determination of the sparticle mass. Background distribution is plotted non-cumulatively and is normalized to the number of signal events.}}
\label{fig:gamma-gamma-inv-mass}
\end{figure}

Moreover, the measured energies of the two scattered protons in \textsc{vfd}s could be used to reconstruct the photon-photon invariant mass $W_{\gamma \gamma}$ and to separate various contributions to the signal. For instance, the expected cumulative $W_{\gamma \gamma}$ distributions for LM1 events with two centrally measured leptons and two detected forward protons are illustrated in Fig.~\ref{fig:gamma-gamma-inv-mass}. \\

If large enough integrated luminosity can be collected, the observed $W_{\gamma \gamma}$ distribution could also be used to measure the masses of supersymmetric particles with a precision of a few GeV, by looking at the minimal c.m.s. energy required to produce a pair of \textsc{susy} particles. Another interesting feature is the capability of computing the missing energy $E^{miss}$ by subtracting the detected lepton energies from the measured two-photon c.m.s. energy. For backgrounds, $E^{miss}$ distributions start at zero missing energy, while for the \textsc{susy} signals they would start only at two times the mass of the \textsc{lsp}~\cite{susy_workshop}.  
%
\section{Photoproduction}
\label{GP}

\subsection{Introduction}

The high luminosity and the high c.m.s. energy of photoproduction processes at \textsc{lhc} offer interesting possibilities for the study of electroweak interaction and for searches beyond the Standard Model (\textsc{bsm}) up to TeV scale. This is illustrated by calculating cross sections for various electroweak and \textsc{bsm} reactions together with their irreducible (\textsc{sm}) background processes.\\ 

In $\gamma p$ interactions, when light quarks are taken massless, a singularity arises in t-channel diagrams, when the outgoing quarks are collinear to the incoming photon. This problem is avoided when the resolved photon contribution is used for some choice of factorization scale, in addition to the direct contribution. Another possibility is to include direct processes only and to apply very lose cuts on specific quantities (i.e on transverse momentum of the outgoing quark) acting as a regulator of this singularity. The second approach gives therefore an approximation of the total cross section. A process for which possible high singularities (and therefore also a significant contribution from resolved photon process) might be expected is for example the photoproduction of single $W$ obtained by scattering a photon on a incoming $u$-quark $\gamma u \rightarrow W^+ d$. A cross section of $13.7(2)~$pb for $7~$TeV incoming protons was computed at Leading Order (\textsc{lo}) using the resolved photon contributions as described in~\cite{Spira:1999ja}. 
It compares very well with the cross section of $14~$pb obtained using the direct process only, calculated with \textsc{CalcHEP}, and for a minimal transverse momentum $p_T > 1~$GeV for the outgoing massless quark. This result has been shown to be rather stable with respect to the $p_T$ cut and indicates that the resolved contribution is small. Another indication of the smallness of resolved contribution was obtained by comparing the \textsc{CalcHEP} computed cross section without $p_T$ cut but using different values of $d$-quark mass as regulator. The cross section was ranging from $18 ~$pb for $m_d=0.001 ~$GeV to $12.8 ~$pb for $m_d=5 ~$GeV (the value of $14 ~$pb being obtained for $m_d=0.8 ~$GeV).\\

In the following, only the direct contributions at \textsc{lo} have been calculated and a cut at $1~$GeV on $p_T$ for the outgoing quark (labelled $q'$) has been applied for all processes with a possible singularity in the t-channel diagrams. This is for instance the case for the cross section $\gamma q \rightarrow W H q'$, even if a $p_T$ cut as low as $0.3~$GeV already gives a very stable result. The list of relevant production cross sections, for several $pp\rightarrow(\gamma g/q \rightarrow X)\rightarrow pXY$, with the final state $X$ including top quarks and/or gauge bosons, are given in Tab.~\ref{back} with the corresponding cuts applied at generator level. Cross sections are evaluated using \textsc{mg/me} or \textsc{CalcHEP}.\\

\begin{table}[!ht]
\begin{center}
\begin{tabular}{p{0.7cm} p{1.3cm}||r p{0.7cm} c c}
\hline
\multicolumn{2}{c}{Processes}  &  \multicolumn{2}{c}{$\mathbf{\sigma}$ [pb]}  & Generator &  Cut  \\\hline
$\gamma g \hspace{0.3cm} \rightarrow$ &$t\overline{t}$    & 1.54\hspace{0.15cm} & & \textsc{mg/me} & -\\
$\gamma q \hspace{0.3cm} \rightarrow$ &$Wt$               & 1.01\hspace{0.15cm} & & $\scriptscriptstyle{//}$ & -\\
&$WWWq'$                                                  & 6.04\hspace{0.15cm} & $\times 10^{-3}$ &$\scriptscriptstyle{//}$ & -\\
&$W^+W^-q'$                                               & 0.605               & & $\scriptscriptstyle{//}$ & -\\
&$W\gamma q'$                                             & 0.349               & & $\scriptscriptstyle{//}$& cut 1\\
&$WZq'$                                                   & 0.151               & & $\scriptscriptstyle{//}$ & -\\
&$Wc$                                                     & 11.4\hspace{0.3cm}  & &$\scriptscriptstyle{//}$ & -\\
&$W^+j$                                                   & 28.1\hspace{0.3cm}  & & $\scriptscriptstyle{//}$ & -\\
&$W^-j$                                                   & 25.0\hspace{0.3cm}  & & $\scriptscriptstyle{//}$ & -\\
$\gamma q/g\rightarrow$& $Wjj$                            & 19.2\hspace{0.3cm}  & & $\scriptscriptstyle{//}$ & -\\
&$Wjjj$                                                   & 8.68\hspace{0.15cm} & & $\scriptscriptstyle{//}$ & -\\
$\gamma q \hspace{0.3cm} \rightarrow$& $Zj$               & 2.62\hspace{0.15cm} & & \textsc{CalcHEP} & -\\
$\gamma q/g\rightarrow$& $Zjj$                            & 1.34\hspace{0.15cm} & & \textsc{mg/me} & -\\
&$Zjjj$                                                   & 0.827               & & $\scriptscriptstyle{//}$ & -\\
$\gamma q \hspace{0.3cm} \rightarrow$& $ZZq'$             & 1.73\hspace{0.15cm} &$\times 10^{-3}$ & $\scriptscriptstyle{//}$ & -\\
& $\gamma j$                                              & 25.3\hspace{0.3cm}  & & \textsc{CalcHEP} & cut 2\\
$\gamma q/g\rightarrow$& $\gamma jj$                      & 12.9\hspace{0.3cm}  & & \textsc{mg/me} & cut 2\\
&$\gamma jjj$                                             & 8.48\hspace{0.15cm} & & $\scriptscriptstyle{//}$ & cut 2\\
$\gamma q \hspace{0.3cm} \rightarrow$& $\gamma \gamma q'$ & 30.4\hspace{0.15cm} & $\times 10^{-3}$ & $\scriptscriptstyle{//}$ & cut 2\\\hline
&$Wb\overline{b}q'$                                       & 45.8\hspace{0.3cm}  & $\times 10^{-3}$ & $\scriptscriptstyle{//}$ & cut 3\\
&$W\tau^+\tau^- q'$                                       & 1.62\hspace{0.15cm} & $\times 10^{-3}$ & $\scriptscriptstyle{//}$ & cut 4\\
&$W\ell^+\ell^- q'$                                       & 19.6\hspace{0.3cm}  &$\times 10^{-3}$ & \textsc{CalcHEP} & cut 5\\
&$W\ell^+\ell^- q'$                                       & 4.43\hspace{0.15cm} &$\times 10^{-3}$ & $\scriptscriptstyle{//}$ & cut 4\\\hline
\end{tabular}
\end{center}
\begin{tabular}{l}
\scriptsize{cut 1 : $p_T^{\gamma}>$ 20~GeV,}\\
\scriptsize{cut 2 : $p_T^{\gamma}>$ 20~GeV, $|\eta^{\gamma}|<5$, $\Delta R(\gamma,j)>$0.3 and $\Delta R(\gamma,\gamma)>$0.3}\\
\scriptsize{cut 3 : $M_{b\overline{b}}>80$~GeV,}\\
\scriptsize{cut 4 : $M_{\ell^+ \ell^-}>110$~GeV,}\\
\scriptsize{cut 5 : 10~GeV $<M_{\ell^+ \ell^-}<70$~GeV,}
\end{tabular}
\caption{\small{Cross sections values and generators used for several $pp\rightarrow(\gamma g/q \rightarrow X)\rightarrow pXY$ processes (here $\ell=e,\mu,\tau$ and $j=u,d,s,c,g$). Different cuts are applied at generator level. For all jets, $p_T^{jet}>$~10~GeV, $|\eta^{jet}|<$~5 and $\Delta R(j,j)>$~0.3. No other cut than the regularization cut $p_T >1~$GeV is applied on $q'$.}}
\label{back}
\end{table}

\begin{figure}[!ht]
\begin{center}
\epsfig{file=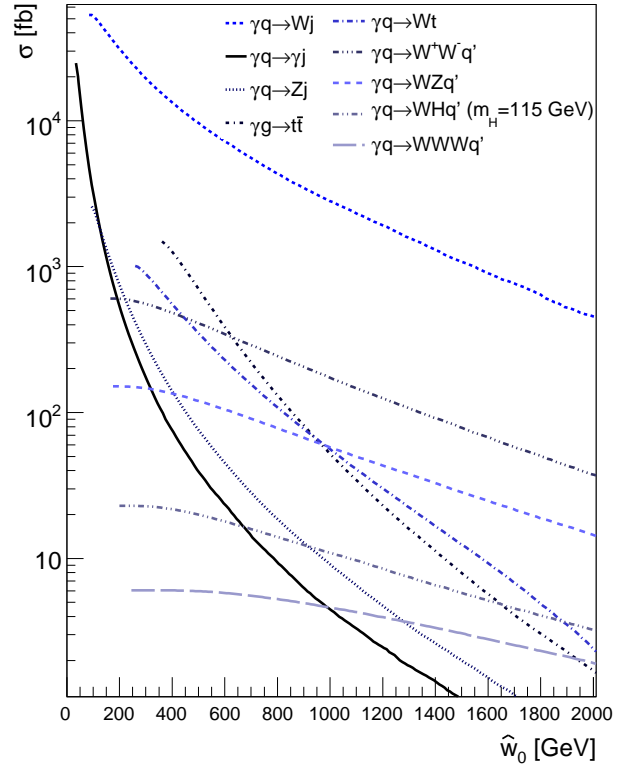,width=\linewidth}
\caption{\small{Cross sections for $pp\rightarrow (\gamma q/g \rightarrow X) \rightarrow pXY$ processes as a function of the minimal photon-parton c.m.s. energy $\hat{W}_0$. Different cuts detailed in Tab.~\ref{back} have been applied.}}
\label{int_gp}
\end{center}
\end{figure}

The Fig.~\ref{int_gp} shows cross sections for $pp\rightarrow pXY$ photoproduction processes as a function of the minimal photon-parton c.m.s. energy. A large variety of processes have sizeable cross section up to the electroweak scale and could therefore be studied during the very low and low luminosity phases of \textsc{lhc}. Interestingly, potentially dangerous Standard Model background processes with hard leptons, missing energy and jets coming from the production of gauge bosons, have cross sections only one or two orders of magnitude higher than those involving top quarks (calculated for $m_t=174.3~$GeV and $m_b=4.7~$GeV).\\

The large top quark photoproduction cross sections ($\cal{O}$1~[pb]) are particularly interesting for measuring top quark related Standard Model parameters, such as the top quark mass and its electric charge. In particular, and in contrast to top quark production in parton-parton interaction, photoproduction of top quark pairs and of single top in association with a $W$ boson have similar cross sections. This will certainly be advantageous in analyses aiming at a measurement of the Cabibbo-Kobayashi-Maskawa (\textsc{ckm}) matrix element $|V_{tb}|$ using the associated $Wt$ production. A more detailed signal-to-noise determination, before and after acceptance cut for single top production, will be discussed in Sec.~\ref{sec.top}\\

\begin{figure}[!ht]
\begin{center}
 \epsfig{file=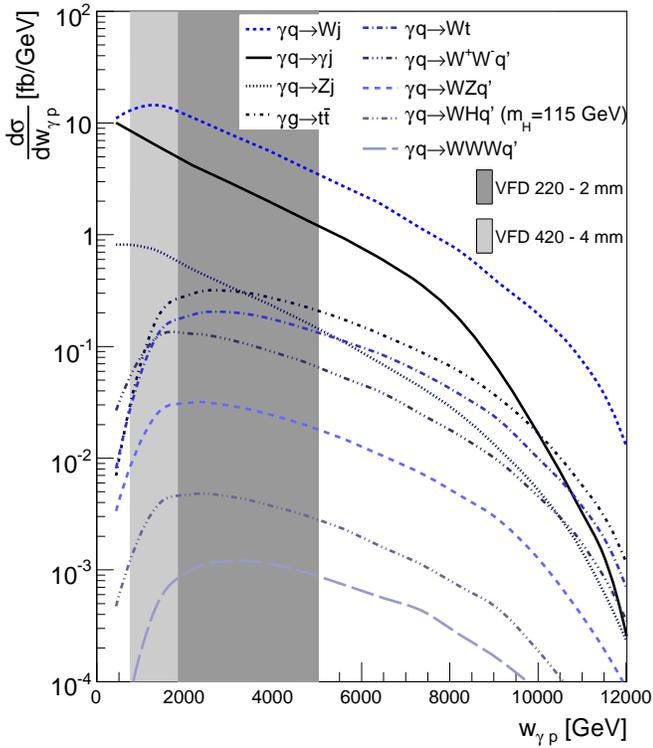,width=\linewidth}
 \caption{\small{Differential cross sections for $pp\rightarrow (\gamma q/g \rightarrow X)\rightarrow pXY$ processes as a function of the photon-proton c.m.s. energy $W_{\gamma p}$. Different cuts detailed in Tab.~\ref{back} have been applied. The acceptance of \textsc{vfd}s (220~m at 2~mm from the beam axis and 420~m at 4~mm from the beam axis) is also sketched. }}
\label{Int}
\end{center}
\end{figure}

The possibility to achieve precise measurements will be enhanced by the capability of tagging photoproduction at higher luminosity. In addition to a large rapidity gap signature in the photon hemisphere, the use of \textsc{vfd}s at 220~m and 420~m will therefore almost certainly be also relevant for photoproduction. The usefulness of using \textsc{vfd}s is illustrated in Fig.~\ref{Int} where differential cross sections for $pp \rightarrow (\gamma q/g \rightarrow X) \rightarrow p X Y$ processes, as a function of the photon-proton c.m.s. energy, is presented together with the acceptance region of \textsc{vfd}s.\\

In order to illustrate the discovery potential of photoproduction at \textsc{lhc}, Sec.~\ref{sec.WH} evaluates the possibility to observe the Standard Model Higgs boson produced in association with a $W$. The associated $WH$ photoproduction cross section reaches more than 20~fb for a 115~GeV Higgs boson and represent more than $2\%$ of the total inclusive $WH$ production at the \textsc{lhc}. Finally, in Sec.~\ref{sec.anotop}, we present the potential for discovery of the anomalous production of single top which could reveal phenomena beyond the Standard Model, and in particular Flavor Changing Neutral Currents (\textsc{fcnc}).\\

In contrast with photon-photon reactions, photoproduction processes discussed in this paper involve topologies with hard jets in the final state. Therefore, in order to take into account the effect of jet algorithms and the efficiency of event selection under realistic experimental conditions, the generated events were passed to \textsc{pythia}~6.227~\cite{pythia} for showering, hadronisation and decay of unstable particles. A fast simulation of a typical \textsc{lhc} multipurpose detector response was then used to define more realistic observables. The fast simulation of the detector's response and the acceptance cuts are described in the following section.\\

\label{sec.GPIntro}
\subsection{Fast detector simulation and acceptance cuts}
\label{sec.DecSimul}

The fast simulation of the detector response is performed with \textsc{delphes}~1.0~\cite{bib:Delphes}. It takes into account geometrical acceptance of sub-detectors and their finite energy resolution, no smearing is applied on particle direction. Charged particles, once are in the fiducial volume of the detector are assumed to be reconstructed with 100$\%$ probability. The energy of each particle produced after hadronization, with a lifetime $c\tau$ bigger than 10~mm is then smeared according to detectors along particule's direction. For particles with a short lifetime such as the $K_s$, the fraction of electromagnetic or hadronic energy is determined according to its decay products. The calorimeter is assumed to cover the pseudorapidity range $|\eta|<3$ and consists in an electromagnetic and an hadronic part. The energy resolution is given by $\sigma_{E}/E=0.05/\sqrt{E} \oplus 0.25/E \oplus 0.0055$ for the electromagnetic part and by $\sigma_{E}/E=0.91/\sqrt{E}\oplus 0.038$ for the hadronic part, where the energy is given in GeV. A very forward calorimeter is assumed to cover $3<|\eta|<5$ with an electromagnetic and hadronic energy resolution function given by $\sigma_{E}/E=1.5/\sqrt{E}\oplus 0.06$ and $\sigma_{E}/E=2.7/\sqrt{E}\oplus 0.13$ respectively.\\

The acceptance cuts applied on leptons and jets used in this section are the following :\\

\begin{itemize}

\item Electrons and muons are reconstructed if they fall into the acceptance of the tracker, assumed to be $|\eta|<2.5$, and have to have a transverse momentum above 10~GeV (the energy resolution of muons is taken to be the same as for electrons). Lepton isolation demands that there is no other charged particles with $p_T>2$~GeV within a cone of $\Delta R<0.5$ around the lepton.\\

\item Jets are reconstructed using a cone algorithm with $R=0.7$ and make only use of the smeared particle momenta. The reconstructed jets are required to have a transverse momentum above 20~GeV and $|\eta|<3.0$. A jet is tagged as $b$-jets if its direction lies in the acceptance of the tracker, $|\eta|<0.5$, and if it is associated to a parent $b$-quark. A $b$-tagging efficiency of $40\%$ is assumed if the jet has a parent $b$-quark. For $c$-jets and light/gluon jets, a fake
$b$-tagging efficiency of $10 \%$ and $1 \%$ respectively is assumed.\\

\item A jet is tagged as a $\tau$-jet if more than $90\%$ of its energy is localized in a cone of $\Delta R=0.15$ around its axis. Moreover, this jet must have its direction in the acceptance of the tracker and have exactly one charged particle with $p_{T}>2$~GeV within a cone $\Delta R<0.4$ around the jet axis. This procedure selects taus decaying hadronically with a typical efficiency of $60\%$. Moreover, the minimal $p_T$ of the $\tau$-jet is required to be 10~GeV.\\

\end{itemize}

\begin{figure}[!ht]
\begin{center}
\epsfig{file=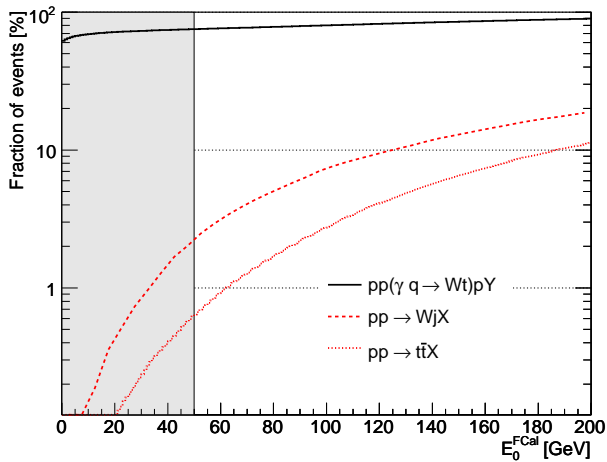,width=\linewidth}
\caption{\small{Fraction of selected events as a function of the rapidity gap cut 
$E^{FCal}_{0}$ displayed for photon-parton induced $Wt$ and parton-parton induced 
$t\overline{t}$ and $Wj$ final states. $E^{FCal}_{0}$ is defined as the cut on the 
minimal of energies $E^{FCal}_{min}$ measured in the two forward calorimeters assumed 
to be located at $3<|\eta|<5$. No other acceptance cut is applied. }}
\label{rapgap}
\end{center}
\end{figure}

Only topologies with at least one high $p_T$ lepton will be studied for obvious reasons related to the capacity of the experiments to trigger events. In addition, during the phase of very low luminosity (i.e.~$10^{32}$ cm$^{-2}$ s$^{-1}$), the pile-up of events is negligible. It is therefore possible to refine the selection of $\gamma$ induced processes, by requiring no significant energy in at least one of the forward calorimeters. For events with $Wt$ in the final state, this simple way of selecting photoproduction has a very good efficiency for an upper cut of 50~GeV as illustrated in Fig.~\ref{rapgap}. This rapidity gap cut will be applied to all processes presented in this section and is referred as $E^{FCal}_{min}$.\\

\subsection{Single top quark photoproduction}
\label{sec.top}

\begin{figure}[!ht]
\begin{center}
\epsfig{file=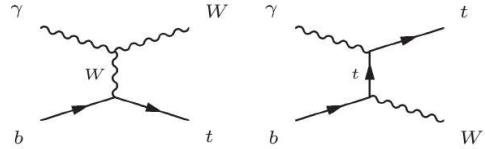,width=0.75\linewidth}
\caption{\small{Leading order Feynman diagrams for associated $W$ and single top 
photoproduction.}}\label{top}
\end{center}
\end{figure}

Photoproduction of single top is dominated by t-channel amplitudes when the top quark is produced in association with a $W$ boson (Fig.~\ref{top}). In contrast to parton-parton interactions where the ratio of $Wt$ associated production cross section to the sum of all top production cross sections is only about $5\%$, it is about 10 times higher in photoproduction. This provides a unique opportunity to study this reaction at the start phase of \textsc{lhc}. While the overall photoproduction of top quark is sensitive to the, yet unmeasured, top quark electrical charge, the $Wt$ associated photoproduction amplitudes are all proportional to the \textsc{ckm} matrix element $|V_{tb}|$.\\

The $\gamma p \rightarrow Wt$ process results in a final state of two on-shell $W$ bosons and a $b$-quark. The studied topologies are (a) $\ell bjj$ for the semi-leptonic decay of the two $W$ bosons and (b) $\ell \ell b$ for the di-leptonic decay. The dominant irreducible background of both channels is expected to stem from the $t\overline{t}$ production, where a jet is not identified. Other $\gamma p$ backgrounds are $Wjjj$ and $WWq'$ processes. The acceptance cuts applied to select $Wt$ events are summarized in Tab.~\ref{CutTop}.\\

\begin{table}[!ht]
\begin{center}
\begin{tabular}{c||c| c}
\hline Acceptance cut &       $\ell bjj$      &  $\ell \ell b$ \\\hline
$\mathrm{N_{\ell}}$           & 1             & 2              \\
$\mathrm{N_{jet}}$            & 2 + 1 $b$-tag & 1 $b$-tag      \\
$\mathrm{|\eta^{jet}_{max}|}$ & 3             & 2.5            \\
\hline
\end{tabular}
\end{center}
\caption{\small{Acceptance cuts for the semi-leptonic and the di-leptonic topologies resulting from $pp\rightarrow(\gamma q \rightarrow Wt)\rightarrow pWtY$ process.}}\label{CutTop}
\end{table}

Tab.~\ref{VisTop} illustrates the signal and background cross sections for two topologies. First, the photoproduction cross section $pp\rightarrow WtYp$ times the branching ratio into the desired topology. Second, the cross sections after acceptance cuts for the signal and the irreducible backgrounds. The inclusive single top cross section after acceptance cuts of 40~fb, with a signal over irreducible background close to 0.6, suggests an easy discovery of this production mechanism with an integrated luminosity of about 1~fb$^{-1}$. Furthermore, a reduction of the background can easily be obtained by adding more specific analysis cuts. However, a more detailed study would be required to also take into account a reducible photoproduction backgrounds and inclusive $pp$ interactions.\\

\begin{table}[!ht]
\begin{center}
\begin{tabular}{l||c| c}
\hline Cross section [fb] & $\ell bjj$ &  $\ell \ell b$   \\\hline
$\sigma \hspace{0.5cm} Wt$                & 440   & 104.3 \\
$\sigma_{acc}$                            & 35.2  & 8.72  \\\hline
\multicolumn{3}{c}{Irreducible background processes}\\\hline
$\sigma_{acc}$ $t\overline{t}$            & 50.01 & 2.98  \\
$\hspace{0.7cm}Wjjj$                      & 17.75 & -     \\
$\hspace{0.7cm}Wb\overline{b}q'$          & 1.06  & -     \\
$\hspace{0.7cm}W^+W^-q'$                  &  -    & 0.18  \\\hline
$\sigma_{acc}$ total                      & 68.82 & 3.16  \\\hline
\end{tabular}
\end{center}
\caption{Cross sections for two $Wt$ induced final states before and after acceptance cuts together with the cross sections of irreducible background processes after acceptance cuts.}\label{VisTop}
\end{table}

\subsection{Associated WH photoproduction}
\label{sec.WH}

The search for $WH$ associated production at \textsc{lhc} will be challenging due to the high $W$+jets, $t\overline{t}$ and $WZ$ cross sections. Indeed, although Standard Model cross sections for the process $pp \rightarrow WHX$ range from 1.5~pb to 425~fb for Higgs boson masses of 115~GeV and 170~GeV respectively\footnote{Cross sections have been calculated with \textsc{mg/me}.}, this reaction is generally not considered as a Higgs discovery channel. However, this production mechanism is sensitive to $WWH$ gauge coupling which might be enhanced when considering models where Higgs boson is fermiophobic. It might also give valuable information on the $Hb\overline{b}$ coupling particularly difficult to determine at the \textsc{lhc}.\\

\begin{figure}[!ht]
\begin{center}
\epsfig{file=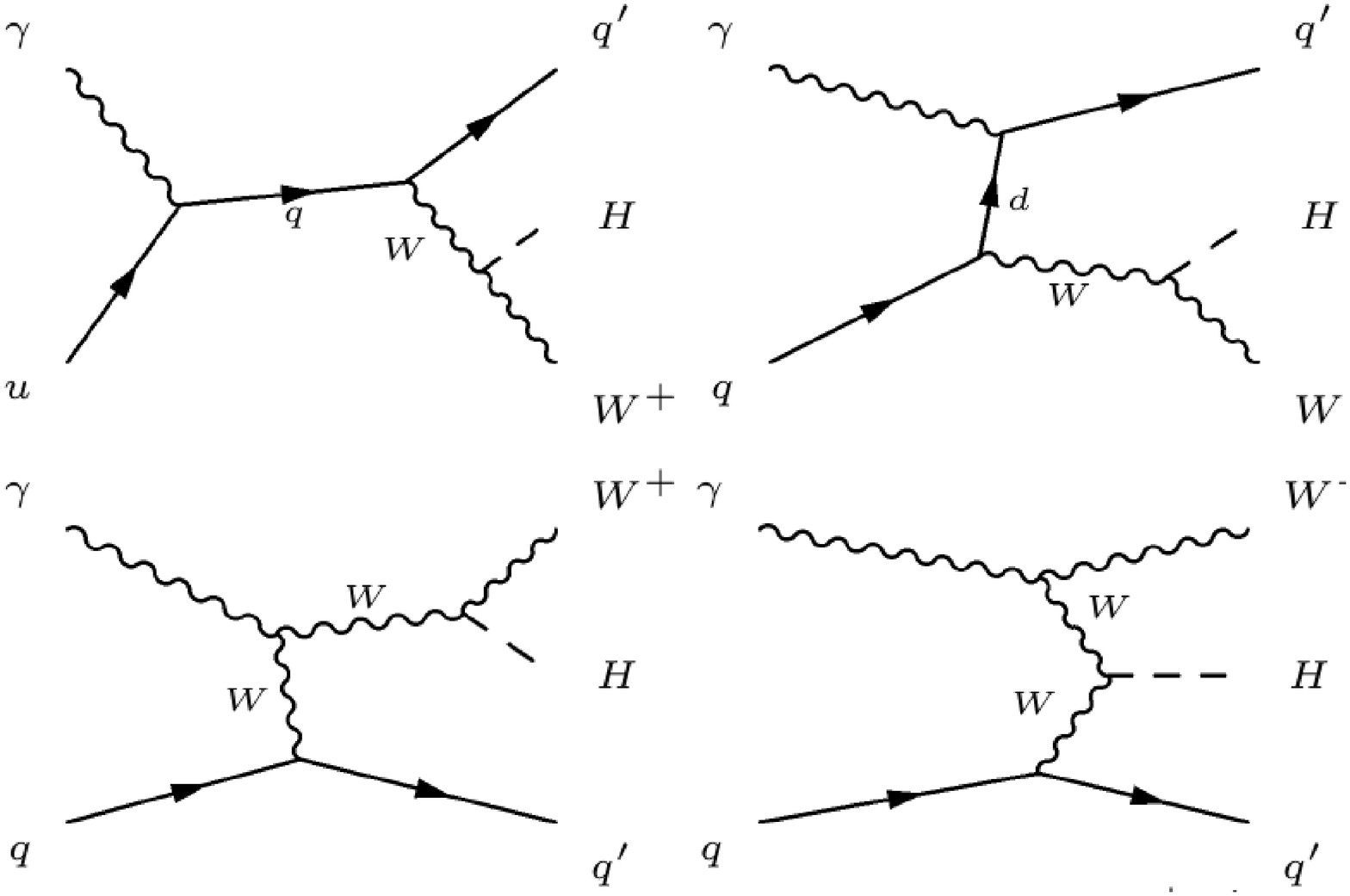,width=0.8\linewidth}
\caption{\small{Leading order Feynman diagrams for associated $WH$ production.}}\label{prod}
\end{center}
\end{figure}

The possibility of using $\gamma p$ collisions to search for $WH$ associate production was already considered at electron-proton colliders~\cite{WH}. At \textsc{lhc} the cross section for $pp\rightarrow(\gamma q \rightarrow WHq')\rightarrow pWHq'Y$ reaches 23~fb for a Higgs boson mass of 115~GeV. It diminishes slowly down to 17.5~fb with increasing Higgs boson masses up to 170~GeV. The dominant Feynman diagrams are shown in Fig.~\ref{prod}. Although cross sections are smaller than the ones induced by initial quarks, the signal to background cross section ratio is improved by more than one order of magnitude as can be seen from Tab.~\ref{back}.\\

Five different topologies have been considered for the signal:
\begin{itemize}
\item $W H \rightarrow \ell \nu b\overline{b}$, $\ell = e, \mu, \tau$
\item $W H \rightarrow W \tau^+ \tau^- \rightarrow jj \ell^+ \ell ^-$, $\ell = e, \mu$
\item $W H \rightarrow W \tau^+ \tau^- \rightarrow jj \ell \tau_h$, $\ell = e, \mu$
\item $W H \rightarrow W W^+ W^- \rightarrow \ell \ell \ell$, $\ell = e, \mu, \tau$
\item $W H \rightarrow W W^+ W^- \rightarrow jj \ell^{\pm} \ell^{\pm}$, $\ell = e, \mu, \tau$. 
\end{itemize}

For each topology, the effects of acceptance cuts as described in Sec.~\ref{sec.DecSimul} and summarized in Tab.~\ref{CutWH} have been evaluated. The reactions considered as irreducible backgrounds are: $t\overline{t}$, $Wt$, $Wb\overline{b}q'$, $W\ell \ell q'$, $WZq'$ and $WWWq'$.\\

\begin{table}[!ht]
\begin{center}
\begin{tabular}{c||c| c| c| c| c}
\hline Acceptance cut & $\ell b\overline{b}$ & $jj \ell \ell$ & $jj \ell \tau_h $ & $\ell \ell \ell$ & $\ell^{\pm} \ell^{\pm} jj$\\\hline
$\mathrm{N_{\ell}}$           & 1         & 2 & 1 & 3       & 2        \\
$\mathrm{N_{\tau_h}}$         & -         & - & 1 & -       & -        \\
$\mathrm{N_{jet}}$            & 2 $b$-tag & 2 & 2 &$\leq 1$ & $\geq 2$ \\
$\mathrm{|\eta^{jet}_{max}|}$ & 3         & 3 & 3 & 3       & 3        \\
\hline
\end{tabular}
\end{center}
\caption{\small{Acceptance cuts for five topologies resulting from $pp \rightarrow pWHq'Y$ photoproduction.}}\label{CutWH}
\end{table}

The first two lines of Tab.~\ref{resWH} show cross sections of the five topologies for Higgs boson masses 115~GeV and 170~GeV before and after the application of acceptance cuts. The second part of the Table presents the expected cross section after acceptance cuts for irreducible background processes.\\

\begin{table}[!ht]
\begin{center}
\begin{tabular}{l||c| c| c| c| c}
\hline Cross section [fb] & $\ell b\overline{b}$ & $jj \ell^{+} 
\ell^{-}$ & $jj \ell \tau_h $ & $\ell \ell \ell$ & $ jj \ell^{\pm} 
\ell^{\pm}$\\\hline
&\multicolumn{3}{c}{$m_H =$ 115~GeV}&\multicolumn{2}{|c}{$m_H 
=$170~GeV}\\\hline
$\sigma \hspace{0.4cm} WHq'$     & 5.42 & 0.14 & 0.52  & 0.55  & 1.17 \\
$\sigma_{acc}$                   & 0.12 & 0.01 & 0.04  & 0.07  & 
0.22\\\hline
\multicolumn{6}{c}{Irreducible background processes}\\\hline
$\sigma_{acc}$ Wt                & 1.13 & 4.24 & 0.98  & -     & -    \\
$\hspace{0.7cm}t\overline{t}$    & 2.43 & 24.5 & 6.34  & -     & -    \\
$\hspace{0.7cm}Wb\overline{b}q'$ & 0.17 & -    & -     & -     & -    \\
$\hspace{0.7cm}W\ell\ell q'$     &-     & 0.41 & 0.08  & 0.43  & 0.11 \\
$\hspace{0.7cm}WZq'$             &-     & 1.47 & 0.12  & 0.98  & 0.06 \\
$\hspace{0.7cm}WWWq'$            &-     & 0.11 & 0.06  & 0.03  & 0.10 \\\hline
$\sigma_{acc}$ total             & 3.73 & 30.8 & 7.56  & 1.44  & 0.28 \\\hline
\end{tabular}
\end{center}
\caption{Cross sections for five $WHq'$ induced final states before and after acceptance cut
together with the cross sections of irreducible background processes after acceptance cuts.}
\label{resWH}
\end{table}

When the decay branching ratio of the Higgs boson into $W$ pair becomes dominant, the same sign lepton signature coming from leptonic decays of two out of the three produced $W$ seems to be promising. It has a signal to noise ratio of about one (which is unique at the \textsc{lhc}) and a luminosity of about 100~fb$^{-1}$ could directly reveal the $HWW$ gauge coupling. More luminosity would probably also allow to probe $Hbb$ coupling for a light Higgs boson, which is known to be very challenging to assess in parton-parton processes.\\  

\subsection{Anomalous top quark photoproduction}
\label{sec.anotop}

In the Standard Model, exclusive single top quark photoproduction at \textsc{lhc} is only possible at a high order of electroweak interactions, since neutral currents preserve quarks flavor at tree level. The observation of a large number of single top events at the \textsc{lhc} would hence be a clean signature of \textsc{fcnc} induced by processes beyond the Standard Model.\\

\begin{figure}[!t]
\begin{center}
\epsfig{file=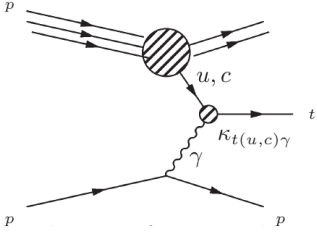,width=0.5\linewidth}
\caption{\small{Leading order Feynman diagram for anomalous single top photoproduction.}}\label{anotop_diag}
\end{center}
\end{figure}

\textsc{fcnc} processes appear in many extensions of the Standard Model, such as two Higgs-doublet models or R-Parity violating supersymmetry. The dominant Feynman diagram contributing to photoproduction of top quarks via \textsc{fcnc}, can be seen in Fig.~\ref{anotop_diag}.\\

The effective lagrangian for this anomalous coupling can be written as~\cite{eff_lag_anotop}: $$ L = iee_t\bar{t}\frac{\sigma_{\mu\nu}q^{\nu}}{\Lambda}k_{tu\gamma}uA^{\mu} + iee_t\bar{t}\frac{\sigma_{\mu\nu}q^{\nu}}{\Lambda}k_{tc\gamma}cA^{\mu} + h.c., $$ where $\sigma^{\mu\nu}$ is defined as $(\gamma^{\mu} \gamma^{\nu} - \gamma^{\nu} \gamma^{\mu})/2$, $q^{\nu}$ being the photon 4-vector and $\Lambda$ an arbitrary scale, conventionally taken as the top mass. The couplings k$_{tu\gamma}$ and k$_{tc\gamma}$ are real and positive such that the cross section takes the form: $$\sigma_{pp \rightarrow tX} = \alpha_u\ k^2_{tu\gamma} + \alpha_c\ k^2_{tc\gamma}. $$ The computed $\alpha$ parameters obtained using \textsc{c}alc\textsc{hep} are the following: $$ \alpha_u = 368\ \texttt{pb} ,\alpha_c = 122\ \texttt{pb}. $$ The best upper limit on $k_{tu\gamma}$ is around 0.14, depending on the top mass~\cite{zeus_st} while the anomalous coupling $k_{tc\gamma}$ has not been probed yet.\\

The final state is composed of a $b$-jet and a $W$ boson. The studied topology is therefore $\ell b$. Main irreducible background processes come from $\gamma p$ interactions producing a $W$ boson and a jet, especially $c$-jets which can be miss-tagged as a $b$-jets. The acceptance cuts applied during the selection of events are summarized in Tab.~\ref{cutanoto}.\\

\begin{table}[!ht]
\begin{center}
\begin{tabular}{c||c}
\hline
Acceptance cut                  & $\ell b$             \\
\hline
$\mathrm{N_{\ell}}$             & 1                    \\
$\mathrm{N_{jet}}$              & 1 $b$-tag            \\
$\mathrm{|\eta^{jet}_{max}|}$   & 2.5                  \\
\hline
\end{tabular}
\end{center}
\caption{\small{Acceptance cuts for the leptonic topology resulting from $pp \rightarrow ptY$ photoproduction process.}}\label{cutanoto}
\end{table}

\begin{figure}[!ht]
\begin{center}
\epsfig{file=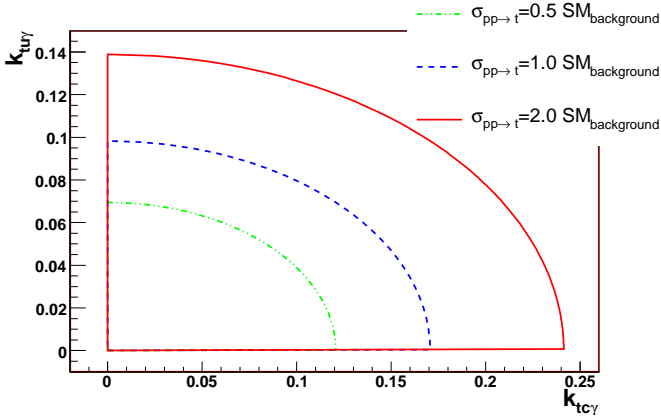,width=\linewidth,clip}
\caption{Constant cross section contours (obtained for $\Lambda = m_{t}$) as a function of anomalous couplings for three different expected signal over irreducible background cross sections after acceptance cuts.}
\label{fig.anomaltop}
\end{center}
\end{figure}

Cross sections after the application of acceptance cuts are shown in Tab.~\ref{xsec_visible}. For the signal, a value of 0.1 was chosen for $k_{tu\gamma}$ while $k_{tc\gamma}$ was set at zero. The resolved $\gamma p \rightarrow Wjq$ has been neglected for the reasons explained in Sec.~\ref{sec.GPIntro}. The discovery potential is illustrated by Fig.~\ref{fig.anomaltop}, showing the line of same cross sections as a function of the anomalous coupling for three different values of cross sections, obtained after acceptance cuts.\\ 

\begin{table}[!ht]
\begin{center}
\begin{tabular}{l||c}
\hline
Cross section [fb]            & $\ell b$      \\
\hline
$\sigma\hspace{0.4cm}$ $t$    & 769.0         \\
$\sigma_{acc}$                & 144.0         \\
\hline
\multicolumn{2}{c}{Irreducible background processes}   \\
\hline
$\sigma_{acc}$ $Wj$           &   56.2        \\
$\hspace{0.7cm}$$Wc$          &   82.8        \\
\hline
$\sigma_{acc}$ total bkg      &  139.0        \\
\hline
\end{tabular}
\caption{Cross sections for the $\ell b$ final state due to anomalous single top photoproduction ($k_{tu\gamma}$ = 0.1,  $k_{tc\gamma}$ = 0) before and after acceptance cut together with the cross sections of irreducible background processes after acceptance cuts.}
\label{xsec_visible}
\end{center}
\end{table}

The present selection together with the assumption that no other background contribution will interfere, would lead to the expectation of a five sigma discovery just below $1$~fb$^{-1}$ of integrated luminosity for anomalous coupling values corresponding to the dashed dotted line of Fig.~\ref{fig.anomaltop}~\cite{jerome}.\\ 
 
However, applying the same statistical treatment as described in section~\ref{sec.WWZZ},
limits on the anomalous couplings k$_{tu\gamma}$ and k$_{tc\gamma}$ can be extracted. These results appear on Tab.~\ref{cl.anotop}. As can be seen, the limits can be greatly improved.\\

\renewcommand{\arraystretch}{1.1}
\begin{table}[!ht]
\centering
\begin{tabular}{c||c|c}
\hline
Coupling        & \multicolumn{2}{c}{Limits}               \\ 
                & L = 1~fb$^{-1}$  & L = 10~fb$^{-1}$      \\
\hline
k$_{tu\gamma}$  & 0.036            & 0.020                 \\
k$_{tc\gamma}$  & 0.062            & 0.035                 \\
\hline
\end{tabular}
\caption{Expected limits for anomalous couplings at 95$\%$ CL for two values of integrated luminosities.}
\label{cl.anotop}
\end{table}
\renewcommand{\arraystretch}{1.}
\subsection{Background}
\label{sec.GPBack}

In this paper, only irreducible photoproduction backgrounds have been considered. However, diffractive and parton-parton processes as well as non irreducible photoproduction backgrounds could also be significant after acceptance cuts.\\

\begin{figure}[!ht]
\begin{center}
\epsfig{file=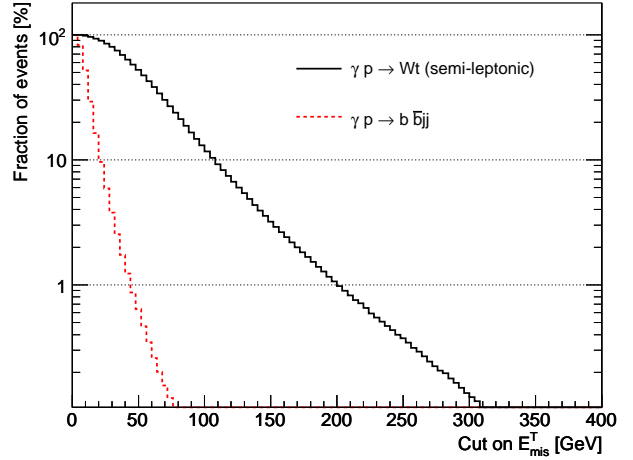,width=\linewidth}
\caption{\small{Fraction of selected events as a function of the transverse missing energy  $E_T^{miss}$ cut displayed for photon-parton induced $Wt$ and the reducible $b\bar{b}jj$ photon-parton induced final state. No other acceptance cut is applied.}}
\label{ptmis}
\end{center}
\end{figure}

Rejection of photoproduction of processes with different final state particles than the signal (called here \textit{reducible background}) is expected to be quite effective when more specific analysis cuts are applied. For instance, no extra cuts on missing energy or on the reconstructed transverse mass of the $W$ boson were required. The rejection power of such a cut is illustrated in Fig.~\ref{ptmis}, where the fraction of events as a function of the transverse missing energy  $E_T^{miss}$ cut is displayed for the photoproduction of $b\overline{b}jj$ and semi-leptonic $Wt$ final states are shown. The $b\overline{b}jj$ sample was generated with \textsc{mg/me} using the same cuts for light jets as those described in Tab.~\ref{back}, leading to a cross section of $91.7$~pb. This final state, which can fake the $Wt$ final state topology when one $b$ quark decays into lepton, can indeed be significantly reduced after applying additional analysis cuts making use of the $E_T^{miss}$ information.\\

Other potentially dangerous backgrounds arise when topologies similar to signal events are produced from the small fraction of parton-parton collisions containing rapidity gaps, but not due to diffractive scattering. As discussed in Sec.~\ref{sec.DecSimul}, a forward rapidity gap condition is in general quite effective. The typical reduction factor obtained using such a condition (energy in one of the forward hemisphere ($3<|\eta|<5$) is smaller than 50~GeV), is $~ 10^{-3}$ and $~ 10^{-2}$ for parton-parton $t\overline{t}$ and $Wj$ production respectively (Fig.~\ref{rapgap}). This reduction factor might not be sufficient for $Wj$ ($j=u,d,s,c,b$) final states given their very large cross section after acceptance cuts : $97~\textrm{pb}$. The rejection can be further improved by tightening the cut which defines the presence of a rapidity gap and also by using other exclusivity conditions related for instance to the number of charged tracks outside jets. Moreover, the $p_T^{jet}$ acceptance cut for jets is also very low and it is probable that any analysis sensitive to jets final states would required harder jets.\\

In order to have an estimate of the rejection power expected on this kind of backgrounds we applied similar acceptance cuts as previously defined in Sec.~\ref{sec.DecSimul}. However, instead of requiring a $p_T^{jet}$ acceptance cut of $20$~GeV we used $30$~GeV. Furthermore, we apply a rapidity gap condition by requiring $E_{min}^{FCal}<$~30~GeV instead of 50~GeV. In addition, an exclusivity condition is applied, requiring no additional track with $p_T>$~0.5~GeV outside a jet cone ($R=0.7$) with $1<\eta<2.5$ in the hemisphere where the rapidity gap is present. With these newly defined acceptance cuts, rapidity gap and exclusivity conditions, efficiency for signal processes drops roughly by a factor of two with respect to acceptance cuts used to obtain numbers of Tab.~\ref{VisTop} and \ref{xsec_visible} while the reduction factors for parton-parton reactions are better than $10^{-3}$.\\

The rejection power is best illustrated in Tab.~\ref{ppback} where representative samples with various hard jets in the final states are shown. A clear dependence on the final topology is observed. Comparing these visible cross sections (after acceptance cuts, rapidity gap and exclusivity conditions), one can see that they are of the same order of magnitude as the corresponding signal and irreducible photo-produced background (Tab.~\ref{VisTop} and \ref{xsec_visible}). For example, semi-leptonic $t\bar{{t}}$ cross section is reduced to $11.8$~fb after these cuts (see Tab.~\ref{ppback}), which is similar to the cross section for the photoproduction of semi-leptonic $t\bar{{t}}$ of typically $46.4$~fb (see Tab.~\ref{VisTop}).\\

\begin{table}[!ht]
\begin{center}
\begin{tabular}{l||ccc}
\hline
Cross section [fb]                      & $pp \rightarrow t\bar{{t}X}$ & $pp \rightarrow tjX$  & $pp \rightarrow WjX$      \\ \hline
$\mathbf{\sigma}$                       & 328 $\times10^3$            & 66.6$\times10^3$     & 86.2$\times10^6$         \\
$\mathbf{\sigma_{acc}}$                 & 25.9$\times10^3$            & 0.96$\times10^3$     & 27.7$\times10^3$         \\
$\mathbf{\sigma_{acc+gap}^{soft}}$      & 155.9                       & 16.0                 & 921.5                    \\
$\mathbf{\sigma_{acc+gap}^{hard}}$      & 71.6                        & 4.00                 & 227.4                    \\
$\mathbf{\sigma_{acc+gap+excl}^{hard}}$ & 19.4                        & 0.67                 & 36.3                     \\ \hline
\end{tabular}
\caption{\small{Cross-sections before and after acceptance cuts corresponding to $l\nu bjj$ topology (for $t\bar{t}$ and $tb$ processes), and to $l\nu b$ (for $Wj$) with $j = u, d, s, c, g$. Acceptance cuts require three (or one in case of $Wj$) jets with $p_T^{jet} > 30~\textrm{GeV}$, a lepton with $p_T^{\ell} >  10~\textrm{GeV}$ and 1 $b$-tagged jet. The rapidity gap condition is $E_{min}^{FCal}<$ 30~GeV, and the additional exclusivity condition requires no track with $p_T>$ 0.5~GeV outside a jet cone ($R=0.7$) with $1<\eta<2.5$. The $t\bar{{t}}$ has semi-leptonic topology and $Wj$ and $tb$ contain one lepton arising from the $W$ decay.}}
\label{ppback}
\end{center}
\end{table}

Providing good control of the energy scale of forwards calorimeters and capability of tagging efficiently rapidity gaps, one expects inclusive parton-parton processes to be negligible at low luminosity or, at most, of the same order of magnitude than the irreducible backgrounds considered in this paper. At high luminosity, however, these backgrounds will remain low, only by means of forward proton tagging capabilities and using more sophisticated exclusivity conditions.\\

Finally, backgrounds due to single diffraction and discussed in Sec.~\ref{sec.bkg.gg} might be of similar size, or even bigger than the ones due to parton-parton interactions. However, the \textsc{vfd}s might be used to verify the diffractive contribution by measuring the transverse momentum distribution of the scattered protons~\cite{piotr_1}, and exclusivity conditions can reduce it further thanks to the additional particle production in diffraction \cite{jerome}.\\
\section{Summary and perspectives}
\label{sec.Summ}

A survey of several high energy photon-photon and photon-proton interactions at the \textsc{lhc} has been presented in this paper. In particular, large cross sections for production of electroweak gauge bosons have been obtained at leading order, using \textsc{mg/me} and \textsc{c}alc\textsc{hep} simulation programs. These large cross sections, up to several picobarns, show the potential of studying massive electroweakly interacting particles in a complementary way to the usual, parton-parton processes. The photon induced processes are often responsible for a few percent of the total proton-proton cross sections and usually have a much lower \textsc{qcd} backgrounds, thereby offer a potentially interesting phenomenology. \\

Interesting studies and searches can already be performed at the initial integrated luminosity of about one inverse femtobarn. The associated photoproduction of a top quark or a $W$ boson is for instance surprisingly large, offering an unique opportunity to measure the fundamental standard model parameters, such as the top quark charge or the $|V_{tb}|$ element of the quark mixing matrix. Anomalous $\gamma qt$ couplings might also be uniquely revealed by photoproduction. \\

Larger integrated luminosity, of about hundred inverse femtobarns, will open complementary ways to search for \textsc{susy} particles in photon-photon interactions. In addition, large luminosity might help to access important information on the Higgs boson coupling to $b$ quarks and $W$ bosons.\\

For the discussed photon-photon processes backgrounds due to parton-parton collisions are expected to be small. For photoproduction, the size of the backgrounds due to parton-parton interactions depends on the capability of experiments to control the energy scale of forward calorimeters and to apply exclusivity conditions. At high luminosity, the efficient selection of photon induced processes is conditioned by the capacity of experiments to tag forward protons by means of dedicated forward detectors.\\

In this context, \textsc{r\&d} programs such as \textsc{fp420} and other developments related to the instrumentation of forward regions of \textsc{cms} and \textsc{atlas} experiments, are particularly important. On the phenomenology side, the ultimate physics potential will only be established, when a full simulation of detector responses can be made. Detailed analyses cuts are currently ongoing in many of the channels considered throughout this paper. In particular the non irreducible photon and full parton-parton backgrounds will be investigated together with an attempt to quantify the systematic uncertainties related to proton tagging techniques.\\
\section*{Acknowledgements}
The authors would like to thank Valery Khoze, Fabio Maltoni and Otto Nachtmann for fruitful discussions.

\end{document}